\documentclass[lettersize,journal]{IEEEtran}
\usepackage{subfigure}
\usepackage{pifont}
\usepackage{tikz}
\usepackage{amsmath, amssymb, amsfonts}
\usepackage{multirow}
\usepackage{graphicx}
\usepackage{caption}
\usepackage{makecell}
\usepackage{pifont}
\usepackage[linesnumbered, ruled, noend]{algorithm2e}

\usepackage{listings}
\usepackage{xcolor}

\definecolor{codegreen}{rgb}{0,0.6,0}
\definecolor{codegray}{rgb}{0.5,0.5,0.5}
\definecolor{codepurple}{rgb}{0.58,0,0.82}
\definecolor{backcolour}{rgb}{0.95,0.95,0.92}
\definecolor{decoratorcolor}{rgb}{0.5,0,0.99}

\lstdefinelanguage{PythonDecorators}{
    morekeywords={@luffy_moe},          
    sensitive=true,
    morecomment=[l]\#,                  
    morestring=[b]",                    
    morestring=[b]',                    
    morestring=[s]{"""}{"""},           
    morestring=[s]{'''}{'''},           
    literate={@}{{\textcolor{decoratorcolor}{@}}}1,
}

\lstdefinestyle{mypy}{
    backgroundcolor=\color{backcolour},   
    commentstyle=\color{codegreen},
    keywordstyle=\color{magenta},
    numberstyle=\tiny\color{codegray},
    stringstyle=\color{codepurple},
    basicstyle=\ttfamily\footnotesize,
    breakatwhitespace=false,         
    breaklines=true,                 
    captionpos=b,                    
    keepspaces=true,  
    columns=flexible,
    numbersep=5pt,                  
    showspaces=false,                
    showstringspaces=false,
    showtabs=false,                  
    tabsize=2,
    morekeywords={@luffy_moe},       
    keywordstyle=[2]\color{decoratorcolor} 
}

\usepackage{tikz}

\usepackage{hyperref}
\hypersetup{hidelinks}

\begin{document}

% \title{\textsc{Luffy}: Accelerating Sparsely-Activated Model Training via Sequence Migration and Token Condensation}
\title{Communication-Efficient Sparsely-Activated Model Training via Sequence Migration and Token Condensation}

 \author{Fahao~Chen,
         Peng~Li,~\IEEEmembership{Senior Member,~IEEE,}
         Zicong~Hong,
         Zhou~Su,~\IEEEmembership{Senior Member,~IEEE,}
         Song~Guo,~\IEEEmembership{Fellow,~IEEE,}
%         % <-this % stops a space
% % \thanks{This paper was produced by the IEEE Publication Technology Group. They are in Piscataway, NJ.}% <-this % stops a space
% % \thanks{Manuscript received April 19, 2021; revised August 16, 2021.}
\IEEEcompsocitemizethanks{\IEEEcompsocthanksitem 
% % This work was supported in part by Japan Society for the Promotion of Science (JSPS) KAKENHI under Grant 24K02932, in part by Japan Science and Technology Agency (JST) PRESTO under Grant 23828673, in part by ROIS NII Open Collaborative Research under Grant 24S0601, in part by the Key-Area Research and Development Program of Guangdong Province under Grant 2021B0101400003, in part by Hong Kong RGC Research Impact Fund under Grant R5060-19, in part by General Research Fund under Grant 152221/19E, Grant 152203/20E, and Grant 152244/21E, in part by the National Natural Science Foundation of China under Grant 61872310, and in part by Shenzhen Science and Technology Innovation Commission under Grant JCYJ20200109142008673.  
% %Corresponding author: Peng Li.

Fahao Chen is with the University of Aizu, Aizuwakamatsu, Japan (e-mail: d8232101@u-aizu.ac.jp).

Peng Li is with the School of Cyber Science and Engineering, Xi'an Jiaotong University, Xi'an, China (e-mail: pengli@xjtu.edu.cn).

Zicong Hong with the Department of Computing, The Hong Kong Polytechnic University, Hong Kong, China (e-mail: zicong.hong@connect.polyu.hk).

Zhou Su is with the School of Cyber Science and Engineering, Xi'an Jiaotong University, Xi'an, China (e-mail: zhousu@ieee.org).

Song Guo is with the Department of Computer Science and Engineering, The Hong Kong University of Science and Technology, Hong Kong, China (e-mail: songguo@cse.ust.hk).}
}

% The paper headers
% \markboth{Transactions on Networking}%
{}
% {Shell \MakeLowercase{\textit{et al.}}: A Sample Article Using IEEEtran.cls for IEEE Journals}

% \IEEEpubid{0000--0000/00\$00.00~\copyright~2021 IEEE}
% Remember, if you use this you must call \IEEEpubidadjcol in the second
% column for its text to clear the IEEEpubid mark.

\maketitle

\begin{abstract}
Mixture-of-Experts (MoE) is an emerging technique for scaling large models with sparse activation. MoE models are typically trained in a distributed manner with an \textit{expert parallelism} scheme, where experts in each MoE layer are distributed across multiple GPUs. However, the default expert parallelism suffers from the heavy network burden due to the all-to-all intermediate data exchange among GPUs before and after the expert run.
% suffers from a high communication cost due to the all-to-all intermediate data exchange among GPUs before and after experts. 
  Some existing works have proposed to reduce intermediate data exchanges by transferring experts to reduce the network loads, however, which would decrease parallelism level of expert execution and make computation inefficient. The weaknesses of existing works motivate us to explore whether it is possible to reduce inter-GPU traffic while maintaining a high degree of expert parallelism. This paper gives a positive response by presenting \textsc{Luffy}, a communication-efficient distributed MoE training system with two new techniques.
  First, \textsc{Luffy} migrates sequences among GPUs to hide heavy token pulling paths within GPUs and avoid copying experts over GPUs. 
  Second, we propose token condensation that identifies similar tokens and then eliminates redundant transmissions. We implement \textsc{Luffy} based on PyTorch and evaluate its performance on a testbed of 16 V100 GPUs. \textsc{Luffy} system can achieve a speedup of up to 2.73$\times$ compared to state-of-the-art MoE training systems.
\end{abstract}


\section{Introduction}\label{sec:introduction}
The recent success of large-scale language models (LLM), e.g., GPT, has demonstrated that model capability often increases with the growth of model sizes~\cite{vaswani2017attention, devlin2018bert, radford2019language, kaplan2020scaling}, however, which comes with a huge computational cost.
Mixture-of-Experts (MoE) 
% \footnote{This paper focuses on MoE based on Transformer models~\cite{vaswani2017attention}. } 
has been proposed as one of the most popular LLM structures, thanks to its unique sparse activation feature with great promises in reducing computational overhead~\cite{lepikhin2020gshard, fedus2022switch}. 
% which only sparsely activate a few of parameters for processing. 
It decomposes the dense part of the model into multiple \textit{experts}. Input sentences, also referred to as sequences, are divided into tokens as basic processing units. A gate network routes these tokens to only a few experts instead of all experts.

% While alleviating the computation costs, a single MoE layer can still easily exceed the limited GPU memory. 

% \begin{figure}[t]
%     \centering
%     \subfigure[Computation-efficient MoE Training.]{
% 		\begin{minipage}[b]{0.45\textwidth}
% 			\includegraphics[width=1\textwidth]{Figures/moe_1.pdf} 
%         \end{minipage}
% 		\label{fig:basic_idea_1}
%     }
%     \subfigure[Communication-efficient MoE Training.]{
% 		\begin{minipage}[b]{0.45\textwidth}
% 			\includegraphics[width=1\textwidth]{Figures/moe_2.pdf} 
%         \end{minipage}
% 		\label{fig:basic_idea_2}
%     }
%     \subfigure[Comp. Comm.-efficient MoE Training (Ours).]{
% 		\begin{minipage}[b]{0.45\textwidth}
% 			\includegraphics[width=1\textwidth]{Figures/moe_3.pdf} 
%         \end{minipage}
% 		\label{fig:basic_idea_3}
%     }
%     \caption{Comparison between the existing works and our idea. Each rectangle represents a token. Tokens with the same pattern mean that they are similar.}
%     \label{fig:basic_idea}
% \end{figure}

\begin{figure*}[t]
    \centering
    \subfigure[Expert Parallelism.]{
		\begin{minipage}[b]{0.23\textwidth}
			\includegraphics[width=1\textwidth]{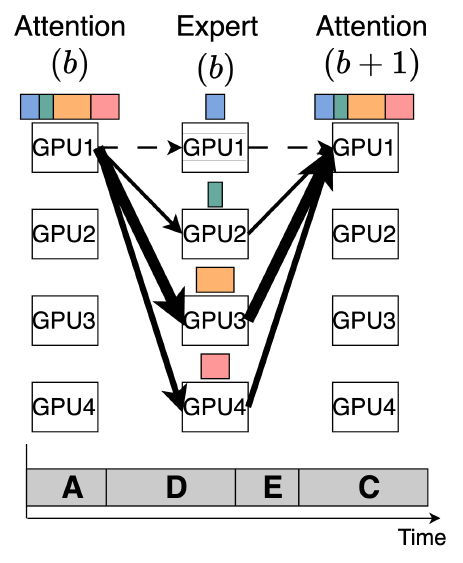} 
        \end{minipage}
		\label{fig:basic_idea_1}
    }
    \subfigure[Expert Transfer.]{
		\begin{minipage}[b]{0.23\textwidth}
			\includegraphics[width=1\textwidth]{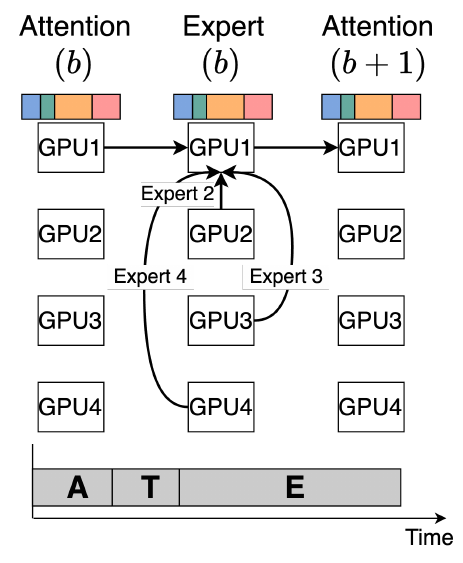} 
        \end{minipage}
		\label{fig:basic_idea_2}
    }
    \subfigure[Sequence Migration (Ours).]{
		\begin{minipage}[b]{0.23\textwidth}
			\includegraphics[width=1\textwidth]{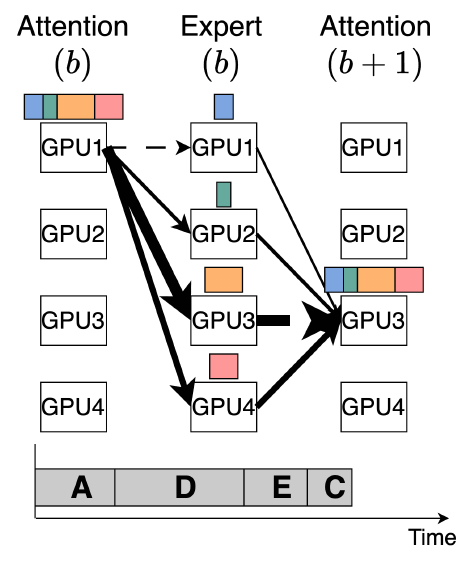} 
        \end{minipage}
		\label{fig:basic_idea_3}
    }
    \subfigure[Token Condensation (Ours).]{
		\begin{minipage}[b]{0.23\textwidth}
			\includegraphics[width=1\textwidth]{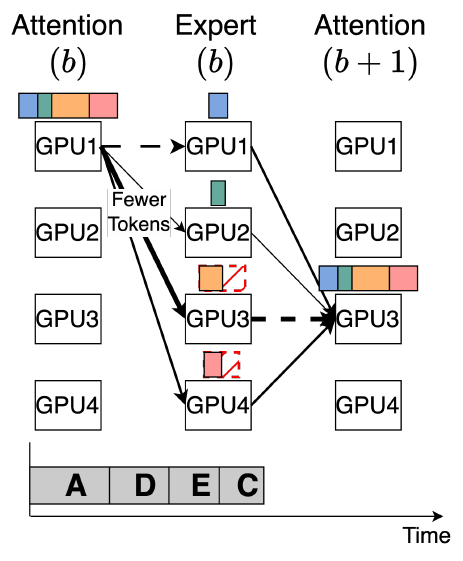} 
        \end{minipage}
		\label{fig:basic_idea_4}
    }
    \caption{Comparison between existing works and our ideas. We assume that each GPU holds one expert (e.g., GPU $k$ holds Expert $k$). \textbf{A}: attention computation; \textbf{E}: expert computation; \textbf{D}: token dispatch; \textbf{C}: token combine; \textbf{T}: expert transfer. Each rectangle represents a part of a sequence, 
    %different colors mean that the part activates different experts, 
    and the width of a rectangle indicates its length. Arrows with solid and dashed lines represent intra-GPU and inter-GPU traffic, respectively, and the arrow width indicates the amount of traffic. The red hatched rectangles in (d) represent the tokens eliminated by token condensation. The block index is denoted by $b$.}
    \label{fig:basic_idea}
\end{figure*}

%Although MoE has achieved great success in various applications, its system-level study still face many open challenges.
Despite the promises of MoE, how to efficiently train MoE models is still an open challenge, mainly because their giant model sizes could easily exceed the memory limit of a single GPU. Therefore, distributed MoE training over multiple GPUs has become one of the hottest topics in AI system research~\cite{rajbhandari2022deepspeed,hwang2023tutel}. Since MoE models have giant sizes and unique structures, some recent works~\cite{lepikhin2020gshard, fedus2022switch} have recognized that traditional parallelism schemes, e.g., data parallelism and model parallelism, can be hardly applied to distributed MoE training. Recently, \emph{expert parallelism}~\cite{lepikhin2020gshard} has been proposed as a novel parallelism scheme dedicated to MoE. As shown in \autoref{fig:basic_idea_1}, experts, which are usually with large sizes, in each MoE layer are distributed across different GPUs, while non-expert components (e.g., multi-head attention) with moderate sizes are replicated over all GPUs. Expert parallelism can maximize GPU resource utilization and thus has become the mainstream of distributed MoE training \cite{nie2023flexmoe, liu2023janus, li2023accelerating}. Several popular open-source model training frameworks, such as Microsoft's DeepSpeed-MoE~\cite{rajbhandari2022deepspeed} and Tutel~\cite{hwang2023tutel}, have supported expert parallelism. 
% \textcolor{red}{In this paper, we mainly focus on efficient MoE training with expert parallelism. Although there have been some attempts to design hybrid parallelism strategies, they are orthogonal to our work.}

However, the default expert parallelism, as shown in \autoref{fig:basic_idea_1}, suffers from a high network burden because of the all-to-all intermediate data exchange among GPUs before and after experts. Many tokens in a GPU may be pushed to experts located at others by the gate network. This process is usually termed as the dispatch phase. After being processed by experts, tokens are pulled back to original GPUs to revert into sequences, in a so-called combine phase. GPUs need to transmit massive tokens through the inter-GPU network in both dispatch and combine phases, which significantly degrades the efficiency of distributed MoE training.
Existing works have revealed that the communication costs become the main bottleneck of the whole system~\cite{liu2022gating, liu2023janus, li2023accelerating}. Some recent works, e.g., Lina~\cite{li2023accelerating}, have proposed to hide this bottleneck by overlapping token transmission and expert computation, but they cannot reduce the size of intermediate data and are still far from fundamentally solving the communication challenge.

Another line of works have proposed to copy remote experts to local GPUs, if there is too much intermediate data pushed out. An example is shown in \autoref{fig:basic_idea_2}. Janus~\cite{liu2023janus} follows this idea and designs sophisticated algorithms to decide when and how to fetch remote experts. FasterMoE~\cite{he2022fastermoe} has proposed a dynamic expert fetching scheme guided by a performance model. However, experts could be large and incur high transmission cost. Moreover, copying more experts to local GPUs intensifies GPU resource competition. The improved communication efficiency is traded with reduced parallelism levels of expert running. In addition, all existing works mainly focus on experts, without much study about multi-head attention, which is the most compute-intensive component of MoE \cite{qu2022dota, you2023vitcod}.

The above facts motivate us to explore whether it is possible to reduce inter-GPU traffic while maintaining a high degree of expert parallelism. This paper gives a positive response by conducting a holistic system study of jointly optimizing communication and computation. We present \textsc{Luffy}, a distributed MoE training system that can significantly improve time-to-accuracy by two key techniques. First, instead of moving experts, \textsc{Luffy} migrates sequences among GPUs to reduce cross-GPU token pulling in the combine phase. As shown in \autoref{fig:basic_idea_3}, if many tokens of a sequence are pushed to a remote GPU, this sequence can be migrated to that GPU. After migration, this sequence is reverted by pulling its tokens to the new location, followed by being fed to the multi-head attention of the next block. Such a sequence migration can hide heavy token pulling paths within GPUs, thus reducing inter-GPU traffic. Meanwhile, it avoids copying experts over GPUs, so that we can save GPU memory and maintain high degree of expert parallelism. 
%Furthermore, sequence migration provides a new chance to optimize attention computation by aligning sequences to reduce padded zeros. 
Furthermore, sequence migration provides a new chance to optimize attention computation by gathering sequences of similar lengths, so that we can reduce padded zeros when aligning them for batch processing. 

We then shift our focus from the combine phase to the dispatch phase, where the second technique, token condensation, can be applied to further reduce inter-GPU traffic.
Token condensation is based on an important observation that a considerable number of tokens pushed to the same expert exhibit high similarity, which has not yet been exploited by existing works. For example, when training the MoE-TransformerXL model, we find that about 62\% tokens pushed to the same expert are very similar, and sending only one of them has almost no influence on the final training accuracy.
Therefore, we propose the token condensation, as shown in \autoref{fig:basic_idea_4}, to identify similar tokens and then to eliminate redundant transmissions. Note that token condensation can reduce traffic not only in the dispatch phase but also in the combine phase because we further find that token similarity can be preserved after passing experts. 

\begin{figure*}[t]
  \begin{minipage}[b]{.3\linewidth}
    \centering
    \includegraphics[width=0.7\linewidth]{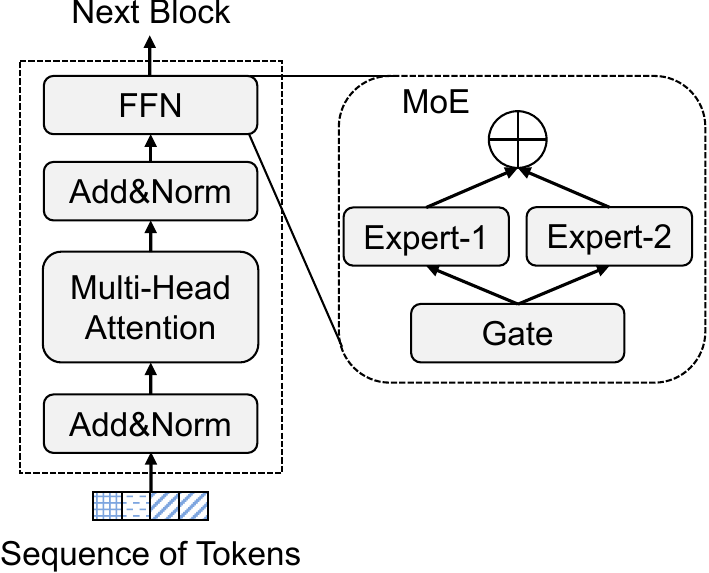}
    \captionof{figure}{An illustration of MoE.}
    \label{fig:moe_back}
  \end{minipage}\hfill
  \begin{minipage}[b]{.6\linewidth}
    \centering
    \resizebox{\linewidth}{!}{
    \begin{tabular}{c|ccc|ccc|ccc}
    \hline
    \multirow{2}{*}{\textbf{Model}} & \multicolumn{3}{c|}{\textbf{Expert=4, Batch = 8}} & \multicolumn{3}{c|}{\textbf{Expert=4, Batch = 16}} & \multicolumn{3}{c}{\textbf{Expert=8, Batch = 8}}\\
    ~ & \textbf{S (GB)} & \textbf{C (ms)} & \textbf{R (\%)} & \textbf{S (GB)} & \textbf{C (ms)} & \textbf{R (\%)} & \textbf{S (GB)} & \textbf{C (ms)} & \textbf{R (\%)}\\
    \hline
    % xl
    MoE-TransformerXL & 3.19 & 327 & 18.1 & 6.15 & 507 & 14.8 & 3.98 & 381 & 30.5\\
    MoE-BERT-Large & 6.73 & 439 & 36.6 & 13.07 & 859 & 40.3 & 7.92 & 477 & 47.5\\
    MoE-GPT2 & 6.53 & 411 & 34.6 & 12.13 & 707 & 35.9 & 7.52 & 452 & 45.9\\
    \hline
    \end{tabular}
    }
        \captionof{table}{Communication bottleneck for distributed MoE training.
    The total number of experts is set equal to the number of GPUs. \textbf{S} means the data transfer size. \textbf{C} and \textbf{R} mean the time of all-to-all communication and its ratio in a batch of training. }
    \label{tab:comm_bottleneck}
  \end{minipage}
\end{figure*}

Although sequence migration and token condensation are promising, it is non-trivial to bring them into full play in practical MoE training due to following technical challenges. \underline{First}, sequence migration needs to decide which sequences should be migrated to which GPUs. \textsc{Luffy} features an algorithm to make migration decisions by jointly considering the token pulling cost and efficiency of the subsequent attention computation. Note that attention computation is affected by two factors. One is about how many sequences should be handled by each GPU, and the other is about whether these sequences have similar lengths, so that we can reduce padded zeros. \underline{Second}, token condensation should have a low overhead in identifying similar tokens. A straightforward pair-wise comparison is with high computational complexity and it can hardly work in practice. We equip \textsc{Luffy} with a fast algorithm to identify similar tokens by fully exploiting token features during MoE training. In addition, token condensation needs to strike a balance between the amount of condensed tokens and training convergence. If more tokens are condensed, we can have lower communication costs but may miss important differences between tokens, which would slow down or even compromise the training convergence. Therefore, \textsc{Luffy} uses a dynamic condensation policy that can adjust token condensation rate according to training convergence status. 

We implement \textsc{Luffy} using PyTorch and evaluate its performance on a testbed consisting of 16 V100 GPUs. The experimental results show that \textsc{Luffy} can achieve a speedup of up to 2.73× compared to state-of-the-art MoE training systems.

% The rest of this paper is organized as follows. Some necessary background about MoE and important motivations are presented in \autoref{background}. A system overview is given in \autoref{sec:overview}, followed by two key techniques, sequence migration and token condensation, in \autoref{sec:migration} and \ref{sec:filtering}, respectively.  \autoref{sec:imple} present details about system implementation. The experimental results and analysis are in \autoref{sec:eva}. Finally, \autoref{sec:conc} concludes this paper.

\section{Background and Motivation}\label{background}
\subsection{MoE and Distributed MoE Training}
Transformer~\cite{vaswani2017attention} emerges as the primary architecture for many complex tasks in natural language processing (NLP), computer vision, and beyond.
As \autoref{fig:moe_back} depicts, the Transformer architecture consists of multiple blocks, each of which includes a multi-head attention layer and a Feed-Forward Network (FFN) layer.
MoE structure has been widely used to scale up Transformer-based models~\cite{riquelme2021scaling, fedus2022switch, zoph2022designing} by replacing the FFN layer with an MoE layer, consisting of a gate network and multiple expert networks that are essentially FFNs.
The instinct behind the MoE is that each expert could be trained to handle a specific kind of input. Thus, the whole model capability becomes stronger as more experts are integrated.

% \subsection{Distributed MoE Training}

However, more experts significantly enlarge the MoE model size, with higher running costs.
For example, Switch Transformer~\cite{fedus2022switch} with 256 experts per block requires more than 50GB memory to accommodate the model parameters. However, the current mainstream GPUs peak at 48GB, while many ranging from 12GB to 24GB.
Thus, traditional data parallelism and pipeline parallelism could hardly be applied for efficient distributed training or inference over multiple GPUs. 
Recently, \textit{expert parallelism}~\cite{fedus2022switch, lepikhin2020gshard} has been proposed as a novel hybrid parallelism dedicated to MoE by exploiting its unique features~\cite{aminabadi2022deepspeed, ma2022bagualu, he2021fastmoe, he2022fastermoe}. 
Specifically, multiple experts are distributed across different GPUs, while other components, such as attention layers, are replicated on GPUs. 
The gate network selects experts for each input token and dispatches them to corresponding GPUs containing the required experts.
This process needs an all-to-all communication between GPUs and is called \emph{dispatch phase}.
After the expert computation, another all-to-all operation is initiated to combine tokens into sequences for the execution of subsequent layers, which is called \emph{combine phase}.
In this paper, we mainly focus on optimizing the expert-parallelism-based distributed MoE training. Although there are some research works about hybrid parallelism strategies~\cite{zhai2023smartmoe, hwang2023tutel}, they are orthogonal to our work.

\subsection{Motivations}
%\subsubsection{Traffic pattern of MoE training}
\noindent\textbf{Traffic pattern of MoE training}:
%We have built a testbed deploy the default expert parallelism
We first conduct experiments to study the communication cost incurred by the all-to-all token push and pull operations in the default expert parallelism\footnote{We focus on the communication cost, which is mainly caused by the all-to-all operations for tokens, excluding gradient synchronization.}. 
% To report their communication costs during training, we train three MoE models, including . 
Table~\ref{tab:comm_bottleneck} shows the data transfer size of all-to-all operations in a training batch of MoE-TransformerXL~\cite{dai2019transformer}, MoE-BERT-Large~\cite{devlin2018bert}, and MoE-GPT2~\cite{radford2019language}. 
We here use top-2 gating and other preliminary experiments follow the same setting.
When training the MoE-BERT-Large with 4 experts per block by setting the batch size as 8, the total data transfer size reaches 6.73GB. Moreover, we can see that the amount of data transmission increases with larger batch sizes and more experts. The communication time becomes a significant portion of the total training time.
When training MoE-BERT-Large with 4 experts per block and a batch size of 8, the communication time is 439ms, which is 36.6\% of the total training time. When the number of experts increases to 8, this ratio grows to 47.5\%. There are similar observations for other models. 

\begin{figure}[t]
\centering
        \subfigure[MoE-TransformerXL.]{
		\begin{minipage}[b]{0.45\textwidth}
			\includegraphics[width=1\textwidth]{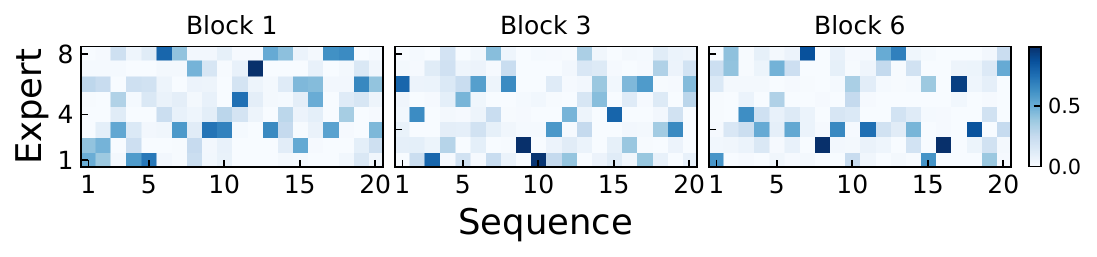} 
  \end{minipage}
		\label{fig:biased_exp_xl}
	}

        \subfigure[MoE-BERT-Large.]{
            \begin{minipage}[b]{0.45\textwidth}
            \includegraphics[width=1\textwidth]{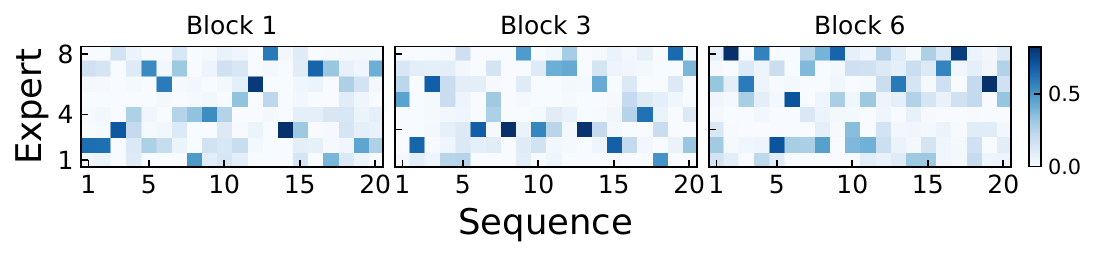}
            \end{minipage}
        \label{fig:biased_exp_bert}
        }
 
        \subfigure[MoE-GPT2.]{
            \begin{minipage}[b]{0.45\textwidth}
            \includegraphics[width=1\textwidth]{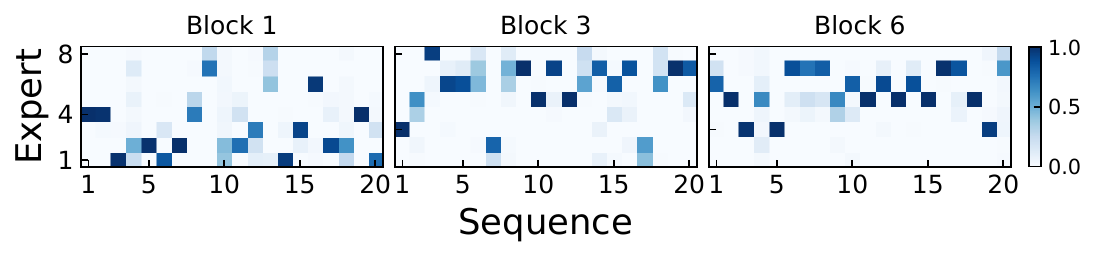}
            \end{minipage}
        \label{fig:biased_exp_gpt}
        }
	\caption{\label{fig:biased_exp}Biased expert activation for sequences under different models after 30 training iterations, where a training iteration indicates the training on a batch of data. Different colors represent hotness values, which indicate the portions of tokens routed to different experts.}
\end{figure}

In addition, we find that each sequence activates very a few experts. We randomly select some sequences and show the portion of tokens pushed to different experts in \autoref{fig:biased_exp}. 
%The statistic information about the number of experts activated by more sequences is plotted in \autoref{fig:biase_dis_.8}. 
Statistically, more than half of sequences use no more than 3 experts when training MoE-TransformerXL and MoE-BERT-Large. MoE-GPT2 has a stronger bias in expert activation, and more than 80\% sequences use only 1 or 2 experts. The biased expert activation is mainly attributed to the data distribution in each sequence, which results in a non-uniform output of the gate network. This characteristic exists throughout the training process, even at the beginning of the training with a randomly initialized gate network. A similar observation has been reported in \cite{he2022fastermoe}.

\begin{figure}[t]
    \centering
    \includegraphics[width=\linewidth]{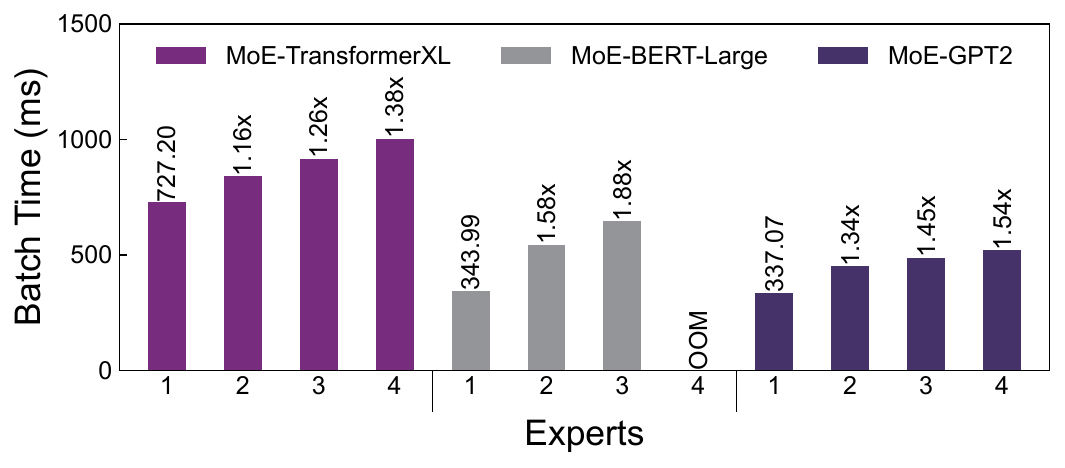}
    \caption{Batch time on one GPU with different number of experts. The batch size is set as 1.}
    \label{fig:comp_motivation}
\end{figure}

\begin{figure}[t]
\centering
        \subfigure[Token similarity over different blocks.]{
		\begin{minipage}[b]{0.45\textwidth}
			\includegraphics[width=\linewidth]{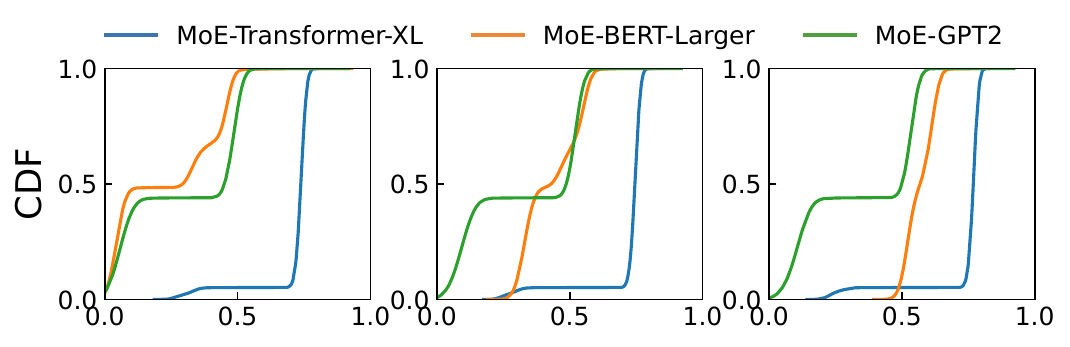} 
  \end{minipage}
		\label{fig:token_sim}
	}

        \subfigure[Token similarity change after the expert execution.]{
            \begin{minipage}[b]{0.45\textwidth}
            \includegraphics[width=\linewidth]{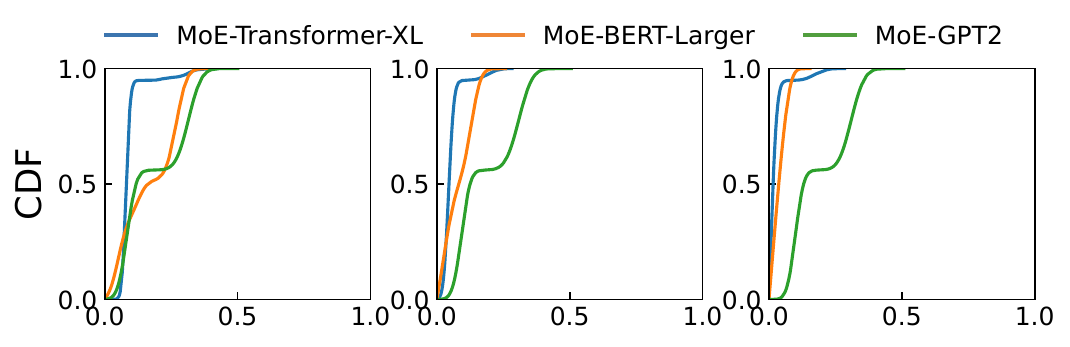}
            \end{minipage}
        \label{fig:token_sim_change}
        }
	\caption{\label{fig:token_sim_all}Token similarity and the change after the expert execution after 30 training iterations. All results are shown over block 1 (\textbf{left}), block 3 (\textbf{middle}), and block 6 (\textbf{right}).}
\end{figure}

To reduce the all-to-all communication cost, some recent works~\cite{liu2023janus,he2022fastermoe,nie2023flexmoe} have proposed to transfer experts instead of intermediate data between GPUs, which are effective when the expert size is smaller than volume of tokens. However, this method cannot fundamentally address the communication bottleneck, especially when facing large expert sizes. For example, recent Mixtral 8$\times$7B \cite{jiang2024mixtral} has 256 experts in total, and each expert is with a size of about 300MB. 
%The worst case of the expert transferring method is exchanging all experts, which incurs significant communication and data loading costs. 
In addition, the execution of experts co-located on the same GPU cannot be well parallelized because of resource competition.
To demonstrate this issue, we migrate different numbers of experts to a single GPU and measure the corresponding running time. As shown in \autoref{fig:comp_motivation}, the computation time increases as more experts are migrated to the same GPU.
For example, as we increase the number of experts from 1 to 3 for MoE-BERT-Large, its expert computation time grows to 1.88$\times$.

\noindent\textbf{Token Similarity}:
%There are a number of similar tokens that are routed to the same experts in the dispatch phase, and their similarity can mostly be preserved after the experts run. 
We collect token embeddings routed to the same experts and compare their cosine similarity, a common metric to evaluate embedding similarity~\cite{antoniak2018evaluating, thongtan2019sentiment, zhou2022problems}. We here use a normalized cosine similarity and its value ranges from [0,1]. A larger value means higher similarity.
Due to space limits, we selectively report the results of three blocks (i.e., block 1, block 3, and block 6) of studied models. 
%where MoE-TransformerXL has 18 blocks, MoE-BERT-Large has 24 blocks, and MoE-GPT2 has 12 blocks in total.
% \textcolor{red}{(how many layers we have in total?)}. 
The results are presented in \autoref{fig:token_sim}, which reveals a significant prevalence of similar tokens across different models. 
For example, about 25\% of token pairs in the first block of the MoE-TransformerXL model have a similarity greater than 0.75. Similarly, in the sixth block of the MoE-BERT-Large model, about 57\% of token pairs have a similarity greater than 0.55. 
In addition, we notice that token pairs tend to show increased similarity in deeper levels. For example, in the first block of the MoE-TransformerXL model, only 25\% of token pairs have a similarity greater than 0.75, but this proportion increases to 85\% in the sixth block. In the case of MoE-GPT2, the proportion of token pairs with similarity above 0.50 increases from 18\% in the first block to 50\% in the sixth block. This trend of higher similarity in deeper layers can be attributed to the significant reduction in the rank of the embedding matrix~\cite{wang2023pangu}.
The above observations suggest a great chance of reducing communication costs by eliminating the transmission of similar tokens.

We further check whether token similarity could be preserved after they pass through the same experts. If similar tokens would be quite different after passing experts, this idea does not work because once we select to transmit only one of similar tokens, their differences are dismissed in the subsequent expert computation. To verify similarity preservation, we select some token pairs whose similarity exceeds a certain threshold, which is set to 0.75 for MoE-TransformerXL, 0.55 for MoE-BERT-Large, and 0.50 for MoE-GPT2, and show their similarity changes in \autoref{fig:token_sim_change}. We can see that the similarity of token pairs has changed only slightly after passing through experts. For example, in the first block of the MoE TransformerXL, about 95\% of token pairs have a similarity change of less than 0.2. Similarly, in the MoE-BERT-Large model, about 36\% of token pairs in the third block have a similarity change below 0.10, and this proportion increases to 98\% in the sixth block.

%-------------------------------------------------------------------------------
\section{System Overview}\label{sec:overview}
%-------------------------------------------------------------------------------

\textsc{Luffy} is designed with the goal of improving time-to-accuracy of distributed MoE training across multiple GPUs. To achieve this goal, \textsc{Luffy} follows several important design principles. \underline{First}, inter-GPU traffic should be minimized. The all-to-all token push and pull operations have been recognized as the main system bottleneck by existing works. The main solution adopted by existing works is to move experts or to exploit network resources with complex scheduling algorithms.
\textsc{Luffy} explores the possibility of fundamentally solving this challenge by reducing number of tokens sent over the network.  
\underline{Second}, the computation part, involving not only experts but also attention layers, of MoE training should also be efficient. Many existing works have excessive concerns about the all-to-all token communication, but share little consideration of expert or attention computation.
\underline{Third}, \textsc{Luffy} should not compromise the training convergence, and thus preserve the quality of the final MoE model. \textsc{Luffy} can allow a certain level of computational approximation during training for acceleration, but approximation errors should be constrained. 

\begin{figure}[t]
    \centering
    \includegraphics[width=\linewidth]{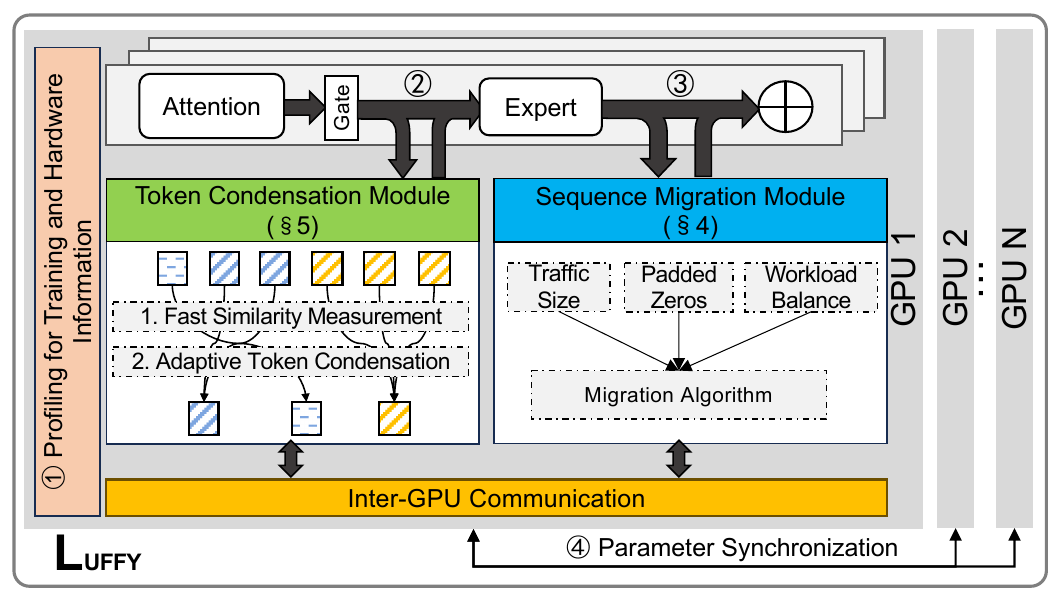}
    \caption{Architecture overview of \textsc{Luffy}.}
    \label{fig:system_overview}
\end{figure}

An overview of \textsc{Luffy}'s design is shown in \autoref{fig:system_overview}. \textsc{Luffy} is based on expert parallelism that each GPU has one or a few experts and a full copy of attention layers. An important design choice that makes \textsc{Luffy} different from existing work is that it does not allow expert movement during training, so that it can parallelize expert running at the maximum level. \textsc{Luffy}'s superiority stems from two novel designs: \textit{sequence migration} and \textit{token condensation}. Sequence migration works in the combine phase, to strategically decide locations where sequences should be re-constructed and then be fed to the subsequent attention layers. An algorithm is designed to make migration decisions by jointly considering token pulling cost and running efficiency of attention layers.
The second key design, token condensation, can be applied for both dispatch and combine phases. It eliminates the transmission of similar tokens to reduce inter-GPU traffic.  Moreover, since only one of similar tokens reaches and goes through the corresponding expert, we can save a lot of expert computation. A fast heuristic algorithm is proposed to quickly identify similar tokens. In addition, the approximation error during expert computation can be constrained to guarantee training convergence. 

The system workflow is as follows. As shown in \autoref{fig:system_overview},
\ding{172} \textsc{Luffy} profiles training information (e.g., batch sizes and sequence lengths) and hardware information (e.g., GPU speed) at the start of each training iteration. Here an iteration is defined as the process of training a batch of data. After profiling, GPUs load training data and proceed to train the MoE model.  
\ding{173} GPUs run attention computations locally, and then send the output to the token condensation module that identifies similar tokens and reduces redundant transmissions. 
\ding{174} Expert computation is launched after receiving tokens. Meanwhile, the sequence migration module decides where tokens should be pulled back for re-construction. After that, \textsc{Luffy} follows these decisions to collect tokens in the combine phase. 
\ding{175} Finally, GPUs compute gradients based on the loss function and synchronize parameters of attention layers and experts for the next iteration.

%-------------------------------------------------------------------------------
\section{Sequence Migration}
\label{sec:migration}
%-------------------------------------------------------------------------------

To minimize token transmissions in the combine phase, an intuitive idea is to migrate each sequence to the GPU accommodating the most of its tokens, so that only a few other ones need to be pushed out and pulled back over the network. However, this approach performs poorly in practice because 
%it would significantly degrades the computational efficiency of subsequent blocks. 
%As shown in \autoref{fig:biased_exp}, many sequences activate experts in a biased way. 
%Thus, if we simply migrate tokens to the GPU that can minimize token transfer for sequence combination, 
some GPUs would be assigned too many sequences, leading to a serious workload imbalance of following attention computation. In addition, if sequences of different lengths are batched together, short sequences must be padded with zeros to align with other long ones, leading to GPU memory waste and additional computation~\cite{fang2021turbotransformers}. In \textsc{Luffy}, we design a sequence migration algorithm that holistically considers the communication cost of the combine phase and the computational efficiency of subsequent blocks. We first present the algorithm design in \autoref{sec:migration_algorithm} and then give details of a cost model that plays an important role in the algorithm in \autoref{sec:performance_modeling}.

\subsection{Migration Algorithm Design}
\label{sec:migration_algorithm}

%%Algorithm-1%%
\begin{algorithm}[t]
\caption{Sequence Migration Algorithm}
\label{alg:migration_alg}
\SetAlgoLined
\KwIn {The set of sequences $N$, the set of GPUs $M$;}
For sequence $i$, we estimate traffic $f_{i,j}$ supposing to migrate $i$ to GPU $j$\;
Put top-$q$ GPUs with minimum traffic into a candidate set $H_{i}$\;
\For{each sequence $i$}{
    \For{each GPU $j\in H_{i}$}{
        $s_{i,j} = T_{att}(B_{j\leftarrow i}, L_{j\leftarrow i}) - T_{att}(B_{j}, L_{j})$\;
    }
    Migrate sequence $i$ to the GPU $j^{*}$ with maximum $s_{i,j}$ if it has sufficient capacity.  
}
\KwRet $j^{*}$ for each sequence\;
% \Ensure $\mathcal{S}$;
% \end{algorithmic}
\end{algorithm}
%%%

The sequence migration problem could be difficult because of dual optimization objectives (i.e., both communication and computation cost) and a large optimization space (i.e., there could be lots of possible combinations of sequences and GPUs). 
To tackle this, we propose a heuristic approach consisting of two steps: one focuses on finding out some candidate GPUs for each sequence with less traffic, and the other step then gathers sequences with similar lengths by migrating them to one of candidate GPUs.

The pseudo codes of the proposed algorithm are shown in \autoref{alg:migration_alg}. For each sequence $i$, we estimate its token pulling traffic if it is migrated to different GPUs. We choose the top-$q$ GPUs with minimum traffic as candidate locations, and include them into set $H_{i}$. A large $q$ can provide more flexibility when we gather sequences with similar lengths in the next step, but may come with higher traffic cost. Setting $q=1$ goes to the other extreme that we put all efforts in minimizing traffic, without considering attention computation.

After getting all $H_{i}$, we continue to find a GPU for each sequence from its candidate set. Similar sequences are expected to stay at the same GPU to reduce padded zeros.
Note that we do not directly count the number of padded zeros as the metric for decision making, as it does not accurately reflect the impact on computational cost. 
Suppose there is a sequence $a$ of length 11 that needs to be migrated. There are two GPUs as candidates, GPU1 with a sequence of length 1, and GPU2 with two sequences of length 6. Migrating sequence $a$ to either GPU results in 10 padded zeros. However, migrating sequence $a$ to GPU2 is a better choice with lower cost because of less attention computation among tokens. 

Therefore, we propose a cost model to estimate the attention computation time. Given $B$ sequences and the maximum sequence length $L$, the cost model can be described as a function $T_{att}(B,L)$, whose details are presented in \autoref{sec:performance_modeling}.
As shown in lines 3-6, for sequence $i$, we select a GPU with the minimum cost growth from its candidate set. Meanwhile, we also consider the capacity constraints of GPUs. A GPU can accommodate more short sequences but less long ones. 
% Meanwhile, we also consider the capacity constraints of GPUs. A GPU can accommodate more short sequences but less long ones. 

\subsection{Cost Model}
\label{sec:performance_modeling}
The performance of the migration algorithm depends on the accurate estimation of attention computation cost. 
As the attention operation is typically compute-bound \cite{nie2023flexmoe},
we can estimate the cost of each attention layer as follows.

\begin{align}
    T_{att}(B,L) = \Bigg\{\underbrace{\frac{3BLd^{2}}{P}}_{\text{Linear projection}}+\underbrace{\frac{2BL^{2}d}{P}}_{\text{Dot-product}}\Bigg\},
\end{align}
where $d$ is the feature dimension and $P$ denotes GPU speed. The rationale of this model is explained as follows. Each attention layer consists of two key operations. First, the input tokens are linearly projected into corresponding queries, keys, and values, denoted by $Q$, $K$, and $V$, respectively. 
Second, a scaled dot product operation is performed to compute the attention scores and weighted outputs as $Attention=softmax(\frac{QK^{T}}{\sqrt{d}})V$.

According to the above workflow, the total operations for each linear projection is $BLd^{2}$. Note that there are three linear projections to generate $Q$, $K$, and $V$, respectively. The scaled dot product operation for each sequence has two parts, where one is the dot product of $Q$ and $K$, resulting in a total operations of $BL^{2}d$. Then, the softmax is applied, followed by the multiplication of $V$ and attention scores, with $BL^{2}d$ operations. Thus, the number of computational operations for the scaled dot product operation is $2BL^{2}d$. 
We ignore the cost of the $softmax$ operation since it is significantly lower than that of matrix multiplication, which has been also verified by \cite{dao2022flashattention}.
The speed $P$ is profiled by running an attention layer several times with varying $B$ and $L$.

\section{Token Condensation} 
\label{sec:filtering}

Token condensation contains two main tasks, measuring token similarity and deciding ``how similar'' are tokens to be condensed. 
An intuitive idea is to compute the pairwise similarity (e.g., cosine similarity) between each pair of tokens and use a predefined similarity threshold to decide which tokens should be condensed. However, this approach can hardly work in practice for the following reasons. 

First, computing pairwise similarity between every two tokens has a high overhead, because of the massive number of tokens involved in MoE training and their high-dimensional embeddings (e.g., each token of MoE-TransformerXL has 1024 dimensions). 
The high overhead would counteract the benefit of token condensation in reducing communication cost. 
To address this challenge, we propose a fast similarity measurement algorithm to efficiently compute the similarity between tokens by strategically skipping unnecessary similarity computations in \autoref{sec:fast_filtering}. 

Second, condensing more tokens means less network traffic, but would increase the risk of compromising convergence due to computational errors. A trade-off between communication efficiency and training convergence should be considered to improve the time-to-accuracy. Moreover, we find that tokens become more similar as the training proceeds, which implies that we can condense more tokens in later training rounds. Motivated by the above observations, we design an adaptive condensation strategy to dynamically set the similarity threshold in \autoref{sec:adaptive}. 

\begin{figure}[t]
\centering
        \subfigure[Similarity change for token pairs in block $b$ and have $s_{b}>0.8$.]{
		\begin{minipage}[b]{0.225\textwidth}
			\includegraphics[width=1\textwidth]{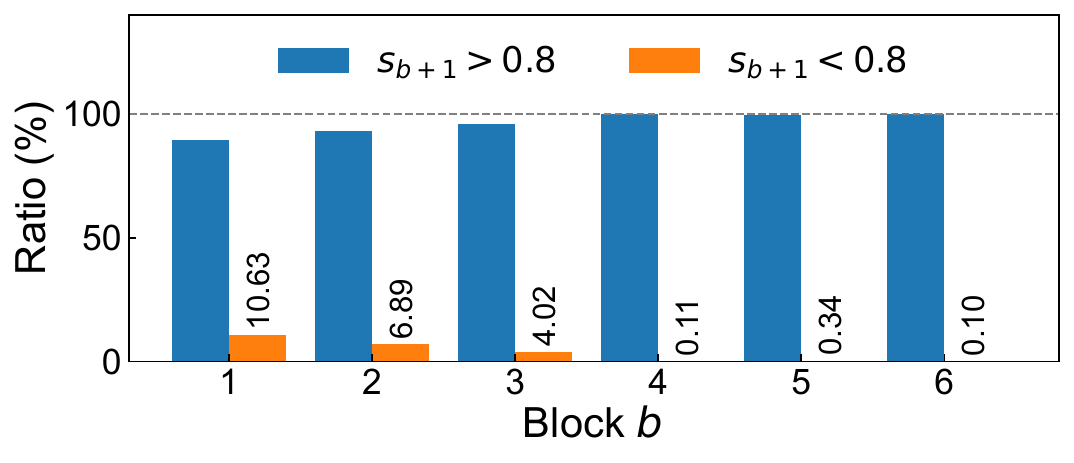} 
  \end{minipage}
		\label{fig:large_keep}
	}
        \subfigure[Similarity change for token pairs in block $b$ and have $s_{b}<0.2$.]{
            \begin{minipage}[b]{0.225\textwidth}
            \includegraphics[width=1\textwidth]{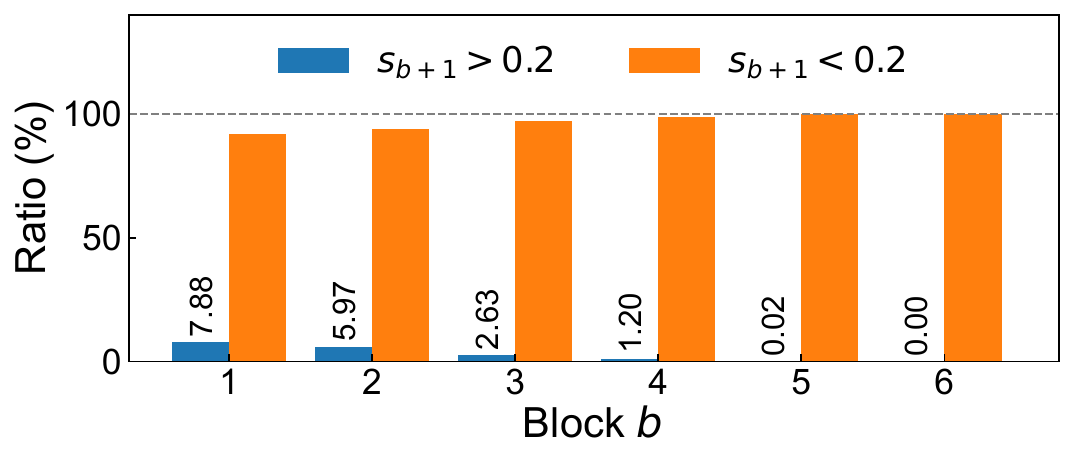}
            \end{minipage}
        \label{fig:small_keep}
        }
	\caption{\label{fig:similarity_keep}Similarity change across consecutive blocks for the MoE-TransformerXL model.}
\end{figure}

\subsection{Fast Similarity Measurement}
\label{sec:fast_filtering}

Our design goal here is to make the measurement process fast by reducing the pairwise similarity computation between tokens. To better present our idea, we model tokens and their similarity relationship as a fully connected graph. 
% An example is shown in \autoref{fig:token_filtering}. 
By default, all edge weights, which represent similarity levels, in the graph need to be measured, leading to high computational overhead. 
However, we find that some tokens are obviously similar or dissimilar according to expert activation and historical knowledge. We can skip their similarity calculation by directly assigning the weights of corresponding edges as 0 or 1.  
Specifically, our fast similarity measurement algorithm works as follows.
% However, we find that some edges can be easily removed, because of obvious unsimilarity obtained from expert activation and historical knowledge. The more edges we remove, the less computation we have for the remaining ones. Specifically, our fast similarity measurement algorithm works as follows.
\begin{enumerate}
    \item \textbf{Measuring similarity by expert activation.} We first find out edges whose associated tokens are pushed to different experts, and set their weights as 0. The design rationale is as follows. Experts of an MoE model are often designed to process different kinds of input. Thus, it is expected that similar tokens are routed to and processed by the same expert. In other words, tokens assigned to different experts are highly unlikely to be similar. Note that tokens going to the same expert are not definitely similar, and we need to continue to check them in the next step.
    \item \textbf{Measuring similarity by historical similarity.} We then identify obviously dissimilar and similar tokens according to historical information. This design is based on the observation that token pairs that are extremely similar or dissimilar in the previous block always tend to keep this pattern. As shown in \autoref{fig:large_keep}, we randomly select some token pairs whose similarity values are greater than 0.8 and check their value changes in several consecutive blocks.
    We can see that about 90\% of token pairs still keep their similarity of greater than 0.8. We also check dissimilar token pairs whose similarity values are less than 0.2 and show results in \autoref{fig:small_keep}. Based on this observation, in the $b$-th block, we identify tokens as similar ones (whose link weights are set to 1) if $s_{b-1}>S_{1}$, or dissimilar ones (whose link weights are 0) if $s_{b-1}<S_{2}$, where $S_{1}$ and $S_{2}$ are two adjustable system parameters. 
    \item \textbf{Calculating similarity of rest token pairs.} Up to now, similarity of most of token pairs has been decided. The rest ones are highly uncertain and we measure them by conducting real cosine similarity calculation. Note that similarity calculations in this step could be quick because they can be easily parallelized.
\end{enumerate}

\subsection{Adaptive Token Condensation}
\label{sec:adaptive}
After token similarity measurement, we need to decide which tokens should be condensed, i.e., not being transmitted in the dispatch phase. Specifically, we delete the edges in the graph whose weights are below a given threshold, which generates a sparse graph composed of multiple subgraphs. For each subgraph, we keep the token with the highest degree for transmission and condense its neighboring tokens. We repeat this process until all tokens are condensed in subgraphs.

% Connected tokens are regarded as similar and are grouped for condensation. 
% For each group of tokens, 
% \textcolor{red}{we keep the token with the highest degree (i.e., maximal token connection) in the dispatch phase and ignore the rest so that the data transmission can be reduced. The reason for choosing the token with the highest degree is to minimize the cumulative similarity error between tokens.}
% we randomly keep one token in the dispatch phase to condense the rest so that the data transmission can be reduced.

To achieve a trade-off between communication efficiency and training convergence, we introduce an adaptive token condensation strategy, which adaptively generates a condensation threshold to ensure the training convergence. The basic idea is as follows. The early stage of MoE training is often accompanied by unstable training convergence. Thus, we need a high threshold to prevent most tokens from being condensed to maintain convergence. 
As the training progresses, the model tends to converge, and we should lower the threshold to condense more tokens so that the data transmission can be reduced as much as possible. 
Specifically, we set the threshold $h_{t}$ for training iteration $t$ according to the loss value in the previous iteration:
\begin{align}
    h_{t} = \frac{1}{1+\exp(l_{norm})},\quad l_{norm} = \frac{l_{ini} - l_{t-1}}{l_{ini}},
\end{align}
where $l_{ini}$ and $l_{t-1}$ are the loss values in the first and previous training iterations, respectively. The normalized loss decrease in the training iteration $t-1$ is denoted by $l_{norm}$. 
The early training stages have a small normalized loss decrease $l_{norm}$, and thus we obtain a large condensation threshold to keep tokens as much as possible. 
As training progresses, $l_{norm}$ becomes larger, indicating that the training tends to be stable. Thus, we lower the threshold $h_t$ to eliminate more tokens. We use an exponential function here to make the threshold unbiased since the loss decrease becomes trivial in the stable stages of the training. 
% The term $C$ is a hyperparameter that determines the tradeoff between training convergence and communication overhead and is set empirically.
% Because tokens become more similar in a deeper block, more tokens have a similarity greater than the threshold in the current training step.
% the ratio of calculated similarities that below a specific similarity reference in the previous block, where the similarity reference may differ from training datasets and models. As we can see, the filtering threshold becomes smaller as the block is deeper, which is because more tokens will become similar. We want to set a small threshold to filter more tokens by keeping more edges in the token graph.

%-------------------------------------------------------------------------------
\section{Implementation}\label{sec:imple}
%-------------------------------------------------------------------------------
\textsc{Luffy} is implemented using PyTorch~\cite{paszke2019pytorch} by adding about 4.5K lines of codes. The developers can easily invoke \textsc{Luffy} as a plug-in-play plugin. 
% \autoref{fig:pythoncode} shows an example about how to use \textsc{Luffy} to train an MoE-based GPT2 model.
% \begin{figure}[t]
% \centering
% \begin{lstlisting}[language=Python]
% import transformers import GPT2LMHeadModel
% model = GPT2LMHeadModel()

% from luffy.function import moe
% moe_model = moe(model, num_experts, top_k)

% # distributed training code...
% \end{lstlisting}
% \caption{An example to illustrate \textsc{Luffy}'s API for MoE implementation.}
% \label{fig:pythoncode}
% \end{figure}

\noindent\textbf{Sequence Migration Controller.} 
In \textsc{Luffy}, a machine is selected to run the sequence migration module and it is called \emph{controller}.
It collects all the information needed for algorithm input and makes migration decisions.
% \textsc{Luffy} deploys the sequence migration controller as an independent process to make migration decisions. 
%The controller runs the migration algorithm (\autoref{sec:migration_algorithm}) with the input of sequence information (e.g., sequence length) and token assignment on GPUs. 
% Since the required information (e.g., sequence locations) used as the algorithm input is distributed across GPUs, the controller adopts a \texttt{torch.distributed.gather()} API to collect them from GPUs. 
The required information used in the migration algorithm, e.g., which GPU each token is dispatched to for expert running, is distributed across GPUs and will be gathered by the controller. This operation can be parallelized with expert running.
To guide the sequence migration, the controller maintains three hash tables: \textit{token\_to\_sequence}, \textit{token\_to\_gpu}, and \textit{sequence\_to\_gpu}, to record the information of execution location for tokens and sequences. For instance, \textit{sequence\_to\_gpu} maintains the information about which GPU each token should be sent to for sequence combining, and it will be updated by the controller according to the migration algorithm outputs. Both tables \textit{token\_to\_sequence} and \textit{sequence\_to\_gpu} will guide GPUs on how to exchange tokens for sequence combining with \texttt{torch.distributed.rpc} APIs. 
% To guide the sequence migration, the controller maintains three hash tables: \textit{token\_to\_sequence}, \textit{token\_to\_gpu}, and \textit{sequence\_to\_gpu}. All tokens, sequences, and GPUs are assigned with unique global IDs. The table \textit{token\_to\_sequence} records the information about which sequence each token belongs to, which is static and generated at the beginning of each iteration of training. This table is also replicated in the memory of each GPU. The table \textit{token\_to\_gpu} represents information about which GPU each token is dispatched to for expert running. This table is updated by each GPU after the gate network runs in each block. The third table \textit{sequence\_to\_gpu} maintains the information about which GPU each token should be sent to for sequence combining. This table is updated by the controller according to the migration algorithm outputs. After the running of gate networks, the migration controller collects the required information (i.e., \textit{token\_to\_gpu}) from GPUs. 
% This operation can be parallelized with expert running. After executing the migration algorithm, the controller updates table \textit{sequence\_to\_gpu} and sends it to all GPUs.
% % using the \texttt{torch.distributed.broadca} \texttt{st()} API. 
% Both tables \textit{token\_to\_sequence} and \textit{sequence\_to\_gpu} can guide GPUs on how to exchange tokens for sequence combining with \texttt{torch.distributed.rpc} APIs. 

\noindent\textbf{Token Condensation Scheduler.} Each GPU maintains a CUDA stream as the token condensation scheduler, which calculates token similarities and conducts token condensation. The scheduler creates a token graph with DGL~\cite{wang2019deep} APIs. Each node represents a token, endowed with two features: the corresponding expert index generated by the gate and the token embedding. Initially, we generate edge features based on the historical similarity between the connected tokens. We define an edge-wise function \texttt{edge\_sim\_calculation} to calculate the similarity between tokens efficiently. To achieve token condensation, the scheduler maintains a hash table \textit{token\_to\_token} to indicate how tokens are condensed. For example, \textit{token\_to\_token}$(i)=j$ represents that token $i$ is condensed and we need to use the expert output of token $j$ to replace it. 
%We randomly sample token IDs with different `value' in \textit{token\_to\_token}. 
% The destination GPU address for token dispatch is generated according to the gate output. 
% When the sequence migration is enabled, the table \textit{token\_to\_token} also needs to be exchanged across GPUs. Each GPU copies received tokens according to the token condensation table and combines them into sequences.

%-------------------------------------------------------------------------------
\section{Evaluation}\label{sec:eva}
%-------------------------------------------------------------------------------

\subsection{Setup}
%-----------------------------------
%%%
\begin{table}[t]
\centering
\resizebox{\linewidth}{!}{
\begin{tabular}{ccccccc}
\hline
Model name & Experts & Layers & $d_{model}$ & $d_{hidden}$ & $len$ & Size\\
\hline
\makecell[c]{MoE-\\Transformer-XL} & \makecell[c]{2, 4\\8, 16} & 18 & 1024 & 4096 & 250 & \makecell[c]{0.44B, 0.74B\\1.34B, 2.55B}\\
\hline
\makecell[c]{MoE-\\BERT-Large} & \makecell[c]{2, 4\\8, 16} & 24 & 768 & 3072 & 512 & \makecell[c]{0.54B, 0.94B\\1.74B, 3.36B}\\
\hline
MoE-GPT2 & \makecell[c]{2, 4\\8, 16} & 12 & 768 & 3072 & 1024 & \makecell[c]{0.18B, 0.29B\\0.52B, 0.97B}\\
\hline
\end{tabular}
}
\caption{Specifications of models for evaluation. $d_{model}$ refers to the dimension of token embedding while $d_{hidden}$ refers to the hidden dimension of FFN layer, i.e., an expert. $B$ is the abbreviation of billion.}
\label{tab:exp_set}
\end{table}
%%%
\noindent\textbf{Testbed. }We evaluate \textsc{Luffy} on a testbed of 16 NVIDIA V100 GPUs with 16GB memory and PCIe connections, aligned with the settings in \cite{he2022fastermoe}. 
% These machines are connected with 10 Gbps ethernet. 
We use Ubuntu 20.04 with Linux kernel version 5.15, NVIDIA driver 525.85, CUDA 11.7, and cuDNN 8.6.0.
% We deploy \textsc{Luffy} under PyTorch~\cite{paszke2019pytorch} of versions 1.12.
% \textcolor{red}{The above settings are aligned with that in \cite{he2022fastermoe}.} 

\noindent\textbf{Models. }We consider three popular MoE models, including (1) MoE-TransformerXL~\cite{dai2019transformer}, a 18-block decoder model; (2) MoE-BERT-Large~\cite{devlin2018bert}, a 24-block encoder model; and (3) MoE-GPT2~\cite{radford2019language}, a 12-block decoder model. 
% \textcolor{red}{All models are pre-trained and we focus on training performance for fine-tuning~\cite{liu2023janus}.}
%To convert original dense models to MoE ones, we replace the FFN layer in each block with the MoE layer. 
The number of experts in each MoE layer varies from 2, 4, 8, to 16. All models are equipped with top-2 gate networks. We set the batch size as 64 for all models. We set the number of experts equal to the number of GPUs, similar to the common practice \cite{fedus2022switch,li2023accelerating}.
% \textcolor{red}{The above evaluation tasks are aligned with that in \cite{he2022fastermoe, liu2023janus, li2023accelerating}.} 
More details of the models used in our experiments are shown in Table \ref{tab:exp_set}.

\noindent\textbf{Baselines. }We compare \textsc{Luffy} with the following baselines. 
(1) Vanilla: the MoE implementation with expert parallelism, adopted by DeepSpeed~\cite{rajbhandari2022deepspeed};
(2) EXT (Expert Transfer): an MoE training paradigm that optimizes the communication cost by transferring activated experts across GPUs, instead of dispatching and combining tokens, which is adopted in Janus~\cite{liu2023janus}; and 
(3) HYT (Hybrid of Token and Expert Transfer): it improves the end-to-end MoE training efficiency by strategically transferring popular experts to all GPUs, which is adopted in FasterMoE~\cite{he2022fastermoe}.
% \textcolor{blue}{
% \noindent\textbf{Baselines. }We compare \textsc{Luffy} with the following baselines. 
% (1) DeepSpeed~\cite{rajbhandari2022deepspeed}: the MoE implementation with expert parallelism and token transfer;
% (2) Janus~\cite{liu2023janus}: an MoE training system that optimizes the communication cost by transferring activated experts across GPUs, instead of dispatching and combining tokens; and 
% (3) FasterMoE~\cite{he2022fastermoe}: it adopts a hybrid of token and expert transfer mechanism to improve the end-to-end MoE training efficiency by strategically transferring popular experts to all GPUs. 
% For a fair comparison, we disable the asynchronous communication mechanism in both Janus and FasterMoE, as it is orthogonal to this work, and \textsc{Luffy} can also benefit from it.
% }

%%
\begin{figure*}[t]
\centering
        \subfigure[MoE-TransformerXL.]{
		\begin{minipage}[b]{0.31\textwidth}
		\includegraphics[width=1\textwidth]{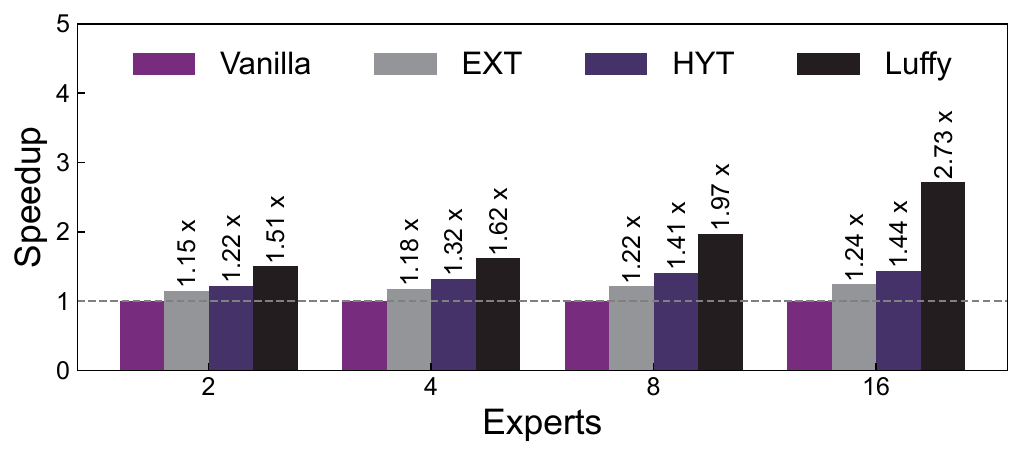} 
            \end{minipage}
		\label{fig:overall_xl}
	}
        \subfigure[MoE-BERT-Large.]{
            \begin{minipage}[b]{0.31\textwidth}
            \includegraphics[width=1\textwidth]{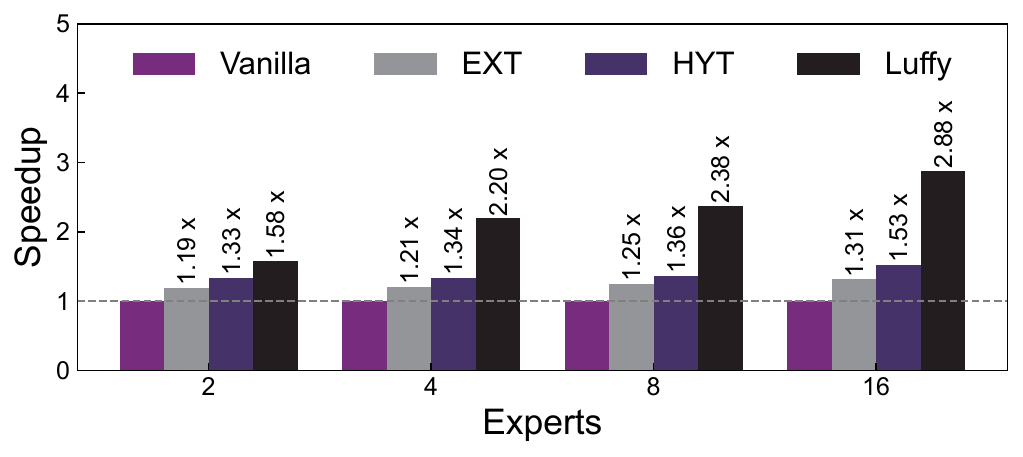}
            \end{minipage}
        \label{fig:overall_bert}
        }
        \subfigure[MoE-GPT2.]{
            \begin{minipage}[b]{0.31\textwidth}
            \includegraphics[width=1\textwidth]{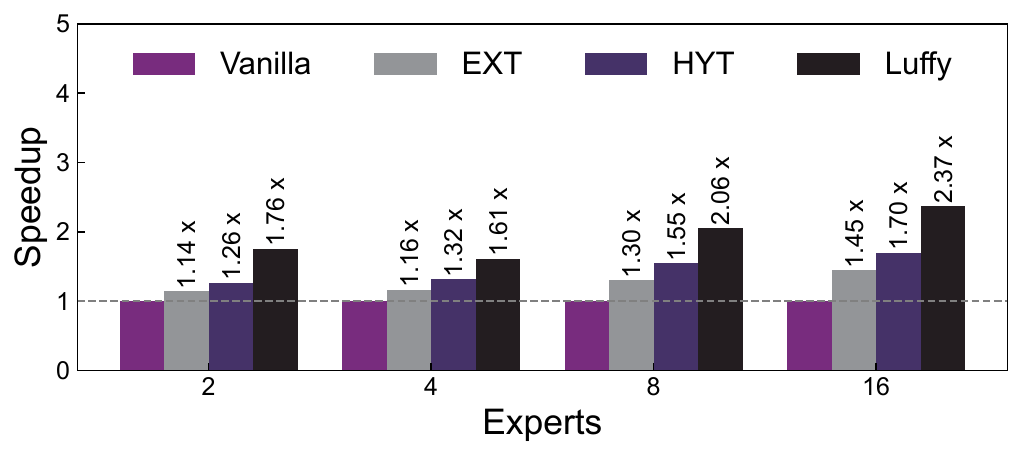}
            \end{minipage}
        \label{fig:overall_gpt}
        }
	\caption{\label{fig:overall_performance}Batch training time speedup of different MoE training systems.}
\end{figure*}
%%
%%%
\begin{table*}[t]
\centering
\resizebox{\linewidth}{!}{
\begin{tabular}{c|c|cc|cc|cc|cc}
\hline
\multirow{2}{*}{\textbf{Model}} & \multirow{2}{*}{\textbf{Method}} & \multicolumn{2}{c|}{\textbf{\#Experts=2}} & \multicolumn{2}{c|}{\textbf{\#Experts=4}} & \multicolumn{2}{c|}{\textbf{\#Experts=8}} & \multicolumn{2}{c}{\textbf{\#Experts=16}}\\
~ & ~ & \textbf{Computation} & \textbf{Communication} & \textbf{Computation} & \textbf{Communication} & \textbf{Computation} & \textbf{Communication} & \textbf{Computation} & \textbf{Communication}\\
\hline
% xl
\multirow{4}{*}{\makecell[c]{MoE-\\TransformerXL}} & \textbf{Vanilla} & 2169 & 843 & 2102 & 1522 & 1923 & 2548 & 1533 & 4599\\
~ & EXT & 2403(\textcolor{red}{0.92$\times$}) & 209(\textcolor{blue}{4.03$\times$}) & 2714(\textcolor{red}{0.78$\times$}) & 370(\textcolor{blue}{4.11$\times$}) & 3054(\textcolor{red}{0.63$\times$}) & 625(\textcolor{blue}{4.07$\times$}) & 3699(\textcolor{red}{0.41$\times$}) & 1233(\textcolor{blue}{3.73$\times$})\\
~ & HYT & 2265(\textcolor{red}{0.96$\times$}) & 197(\textcolor{blue}{4.28$\times$}) & 2387(\textcolor{red}{0.84$\times$}) & 357(\textcolor{blue}{4.26$\times$}) & 2629(\textcolor{red}{0.72$\times$}) & 539(\textcolor{blue}{4.73$\times$}) & 3204(\textcolor{red}{0.48$\times$}) & 1068(\textcolor{blue}{4.31$\times$})\\
~ & \textsc{Luffy} & 1521(\textcolor{blue}{1.43$\times$}) & 480(\textcolor{blue}{1.76$\times$})& 1389(\textcolor{blue}{1.51$\times$}) & 851(\textcolor{blue}{1.79$\times$}) & 1225(\textcolor{blue}{1.57$\times$}) & 1043(\textcolor{blue}{2.35$\times$}) & 1012(\textcolor{blue}{1.52$\times$}) & 1238(\textcolor{blue}{3.72$\times$})\\
\hline
% bert
\multirow{4}{*}{\makecell[c]{MoE-\\BERT-Large}} & \textbf{Vanilla} & 973 & 899 & 953 & 2122 & 918 & 3072 & 756 & 4284\\
~ & EXT & 1258(\textcolor{red}{0.77$\times$}) & 314(\textcolor{blue}{2.87$\times$})& 1989(\textcolor{red}{0.48$\times$}) & 561(\textcolor{blue}{3.60$\times$}) & 2011(\textcolor{red}{0.45$\times$}) & 1181(\textcolor{blue}{2.60$\times$})& 2112(\textcolor{red}{0.36$\times$}) & 1728(\textcolor{blue}{2.48$\times$})\\
~ & HYT & 1123(\textcolor{red}{0.87$\times$}) & 281(\textcolor{blue}{3.21$\times$}) & 1794(\textcolor{red}{0.53$\times$}) & 506(\textcolor{blue}{3.99$\times$}) & 1843(\textcolor{red}{0.49$\times$}) & 1083(\textcolor{blue}{2.84$\times$}) & 1914(\textcolor{red}{0.39$\times$}) & 1386(\textcolor{blue}{3.09$\times$})\\
~ & \textsc{Luffy} & 784(\textcolor{blue}{1.24$\times$}) & 404(\textcolor{blue}{2.23$\times$})& 728(\textcolor{blue}{1.31$\times$}) & 672(\textcolor{blue}{3.01$\times$}) & 638(\textcolor{blue}{1.44$\times$}) & 1042(\textcolor{blue}{2.95$\times$}) & 525(\textcolor{blue}{1.44$\times$}) & 1225(\textcolor{blue}{3.49$\times$})\\
\hline
% gpt
\multirow{4}{*}{MoE-GPT2} & \textbf{Vanilla} & 955 & 881 & 847 & 1573 & 774 & 2592 & 676 & 3834\\
~ & EXT & 1399(\textcolor{red}{0.68$\times$}) & 209(\textcolor{blue}{4.22$\times$}) & 1706(\textcolor{red}{0.49$\times$}) & 374(\textcolor{blue}{4.21$\times$}) & 2048(\textcolor{red}{0.38$\times$}) & 544(\textcolor{blue}{4.77$\times$}) & 2402(\textcolor{red}{0.28$\times$}) & 718(\textcolor{blue}{5.34$\times$})\\
~ & HYT & 1278(\textcolor{red}{0.75$\times$}) & 174(\textcolor{blue}{5.06$\times$}) & 1509(\textcolor{red}{0.56$\times$}) & 331(\textcolor{blue}{4.75$\times$}) & 1741(\textcolor{red}{0.45$\times$}) & 435(\textcolor{blue}{5.96$\times$}) & 2095(\textcolor{red}{0.32$\times$}) & 557(\textcolor{blue}{6.88$\times$}) \\
~ & \textsc{Luffy} & 752(\textcolor{blue}{1.27$\times$}) & 292(\textcolor{blue}{3.02$\times$}) & 724(\textcolor{blue}{1.17$\times$}) & 780(\textcolor{blue}{2.02$\times$}) & 669(\textcolor{blue}{1.16$\times$}) & 963(\textcolor{blue}{2.69$\times$}) & 571(\textcolor{blue}{1.18$\times$}) & 1330(\textcolor{blue}{2.88$\times$})\\
\hline
\end{tabular}
}
\caption{Performance breakdown. We use Vanilla as the baseline. \textcolor{blue}{Blue} values indicate efficiency improvements while \textcolor{red}{red} values indicate efficiency degradation.}
\label{tab:break_down}
\end{table*}
%%%
\subsection{End-to-End Performance}
\autoref{fig:overall_performance} shows the overall speedup by normalizing the average iteration time over that of Vanilla, where an iteration indicates the training on a batch of data. All methods use the same configurations when training on the same model, such as sequence length and batch size. \textsc{Luffy} outperforms other baselines under all models and its superiority becomes clearer when there are more experts. For instance, with a number of experts ranging from 4 to 16, \textsc{Luffy} provides a speedup from 1.51$\times$ to 2.73$\times$ over the vanilla on the MoE-TransformerXL model. This is because more experts means more all-to-all traffic, but \textsc{Luffy} has stronger capability to reduce the total number of data transfers through token condensation and sequence migration, resulting in a higher speedup.

EXT and HYT also show obvious improvement over the vanilla solution. However, it's important to note that they may introduce significant competition for GPU resources during model computation, which can compromise the benefits of expert transfer by reducing parallelism. As a result, their performance improvement may be limited. \textsc{Luffy} optimizes training efficiency by jointly considering communication costs and computation efficiency, resulting in higher performance gains. Compared to EXT and HYT, \textsc{Luffy} achieves up to 1.65$\times$ and 1.46$\times$ speedup, respectively. 
More details about the results are as follows.

\noindent\textbf{MoE-TransformerXL.} The MoE-TransformerXL has larger experts, which can result in higher communication cost when transferring them. 
%It's important to balance the communication costs with the need for expert transfer. 
EXT copies remote experts to local GPUs once they are activated, which may not always be the best choice. Thus, it has smaller speedup under MoE-TransformerXL than other models. 
In contrast, HYT and \textsc{Luffy} well consider expert sizes and have achieved higher performance. 
%\textsc{Luffy} can still provide about 1.62$\times$ speedup.

\noindent\textbf{MoE-BERT-Large.} The MoE-BERT-Large model has more experts because of its large number of MoE blocks. When transferring these experts, there may be competition for GPU resources, which can result in longer computation times for expert running. Furthermore, the negative impact of expert transfer is amplified with more MoE blocks. Our \textsc{Luffy} model achieves approximately 1.61$\times$ and 1.80$\times$ speedup compared to EXT and HYT, respectively. 

\noindent\textbf{MoE-GPT2.} The MoE-GPT2 model has a small number of experts but a large number of tokens in each MoE block due to the long sequence length. Therefore, both EXT and HYT provide a high speedup with expert transfer. However, our \textsc{Luffy} still achieves respective 1.33$\times$ and 1.53$\times$ speedups compared to EXT and HYT.

Moreover, the speedup provided by \textsc{Luffy} is from not only communication reduction but also computation savings. This is because token condensation decreases the number of tokens processed by experts, thereby reducing computation costs. We show more results and analysis of communication and computation improvements in \autoref{sec:breakdown}.

\subsection{Performance Breakdown}\label{sec:breakdown}
%We report the average time cost of training three models for one iteration and break it down into computation, dispatch, and combining costs. The average statistics are profiled from a large number of iterations, as shown in \autoref{tab:break_down}. 
We then break down the batch training time into three parts (computation time for attention and expert running, and communication time for token and expert transfer) and show the results in \autoref{tab:break_down}. We report the time cost in milliseconds for each part as well as the speedup, compared to Vanilla. EXT and HYT can significantly reduce the communication cost by about 3.84$\times$ and 4.45$\times$ compared to Vanilla. 
However, they sacrifice the parallelism level of expert running by introducing intensive GPU resource contention. The computation cost is increased by about 1.81$\times$ and 1.62$\times$ with EXT and HYT, respectively. The computation costs of EXT and HYT increase with more experts. For the MoE-GPT2 model with 16 experts, the computation costs of EXT and HYT are 3.57$\times$ and 3.13$\times$ higher than that of Vanilla.
In contrast, our \textsc{Luffy} optimizes communication efficiency without sacrificing the parallelism level of expert running. In addition, the sequence migration and token condensation modules introduced by \textsc{Luffy} can also optimize the efficiency in both expert and attention running computation. As a result, \textsc{Luffy} respectively achieves average speedups of 1.35$\times$ and 2.66$\times$ on communication and computation, compared to Vanilla.

% Furthermore, for Vanilla and our \textsc{Luffy}, the communication cost increases as the number of experts increases. The reason is that more experts result in more token transfers in the dispatch and combine phases. However, our \textsc{Luffy} can reduce the token transfers by exploiting the unique characteristics of the traffic pattern of MoE training and token similarity, significantly reducing the communication costs and achieving an average gain of 2.91$\times$.

\subsection{Ablation Study}
\begin{figure*}[t]
\centering
        \subfigure[MoE-TransformerXL.]{
		\begin{minipage}[b]{0.31\textwidth}
		\includegraphics[width=1\textwidth]{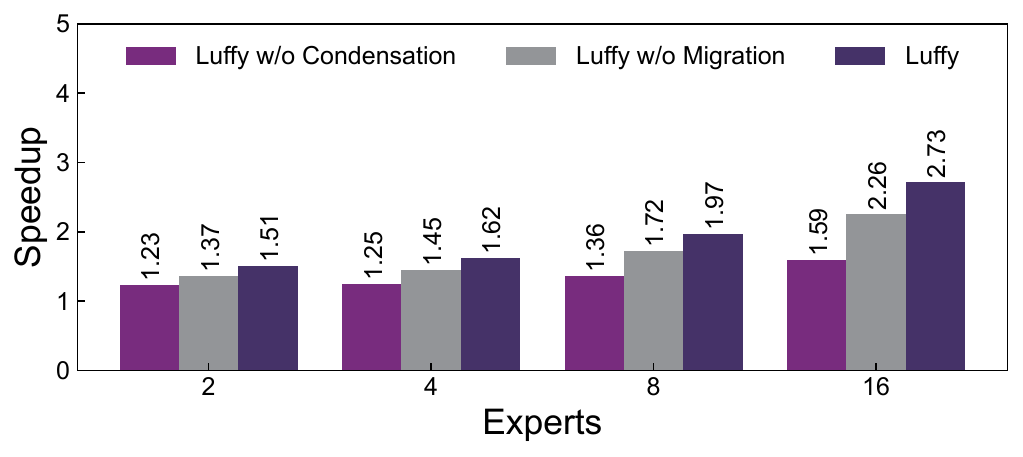} 
            \end{minipage}
		\label{abla_xl}
	}
        \subfigure[MoE-BERT-Large.]{
            \begin{minipage}[b]{0.31\textwidth}
            \includegraphics[width=1\textwidth]{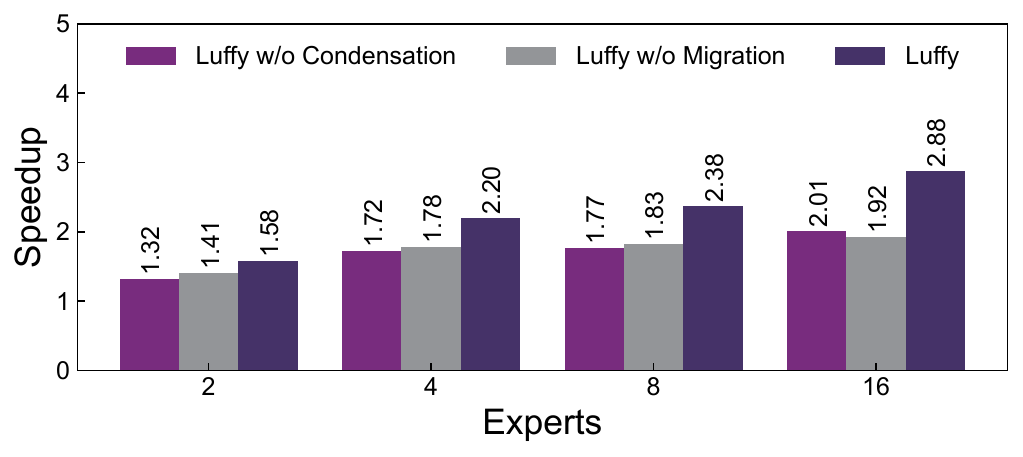}
            \end{minipage}
        \label{abla_bert}
        }
        \subfigure[MoE-GPT2.]{
            \begin{minipage}[b]{0.31\textwidth}
            \includegraphics[width=1\textwidth]{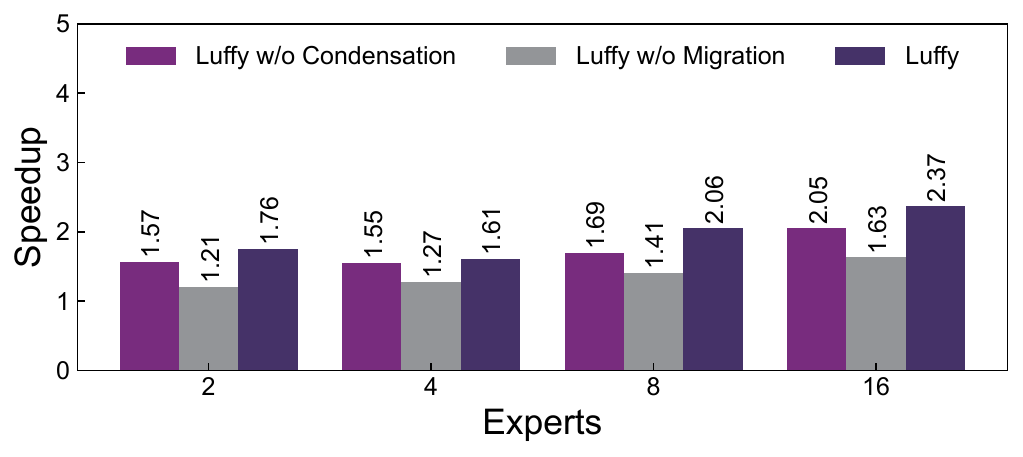}
            \end{minipage}
        \label{abla_gpt}
        }
	\caption{\label{ablation_performance}Performance improvements of the optimizations separately. We use Vanilla as the baseline and show speedups of different optimizations in \textsc{Luffy}.}
\end{figure*}
%%
% %% convergence
% \begin{figure*}[t]
% \centering
%         \subfigure[MoE-TransformerXL.]{
% 		\begin{minipage}[b]{0.31\textwidth}
% 		\includegraphics[width=1\textwidth]{Figures/Experiment/convergence_xl.eps} 
%             \end{minipage}
% 		\label{conv_xl}
% 	}
%         \subfigure[MoE-BERT-Large.]{
%             \begin{minipage}[b]{0.31\textwidth}
%             \includegraphics[width=1\textwidth]{Figures/Experiment/convergence_bert.eps}
%             \end{minipage}
%         \label{conv_bert}
%         }
%         \subfigure[MoE-GPT2.]{
%             \begin{minipage}[b]{0.31\textwidth}
%             \includegraphics[width=1\textwidth]{Figures/Experiment/convergence_gpt.eps}
%             \end{minipage}
%         \label{conv_gpt}
%         }
% 	\caption{\label{fig:convergence}Convergence of \textsc{Luffy} with different similarity thresholds. We use Vanilla as the baseline.}
% \end{figure*}
% %%
We conduct ablation experiments to study the performance improvement of different components in \textsc{Luffy}. We use Vanilla as a baseline and report the average speedup of different components in \autoref{ablation_performance}. We see that the token condensation and sequence migration modules provide different performance gains for different MoE models. Specifically, for the MoE-TransformerXL model, we find that token condensation provides a higher performance improvement compared to sequence migration. This is because MoE-TransformerXL model has more similar tokens (as demonstrated in \autoref{fig:token_sim}) and thus token condensation can condense more tokens, leading to reduce inter-GPU traffic. When only the token condensation is enabled, \textsc{Luffy} provides about 1.74$\times$ speedup over the baseline.
In contrast, in the MoE-GPT2 model, the tokens have less similarity and the benefit of token condensation is less than in the MoE-TransformerXL model, which brings only 1.38$\times$ speedup. 
On the other hand, MoE-GPT2 shows stronger biased expert activation and thus we have a large optimization space that can be exploited by the sequence migration, with about a 1.72$\times$ speedup. In the MoE-BERT-Large model, both token condensation and sequence migration provide high performance gains.

\subsection{Convergence Evaluation}
\begin{table}[t]
\centering
\resizebox{\linewidth}{!}{
\begin{tabular}{c|c|c|c|c|c|c}
\hline
Model name & Dataset & Metric & Vanilla & Luffy ($h=0.3$) & Luffy ($h=0.8$) & \textsc{Luffy}\\
\hline
\makecell[c]{MoE-\\Transformer-XL} & WikiText-103 & PPL $\downarrow$ & \textbf{25.13} & 31.52 & 25.79 & 25.28\\
\hline
\makecell[c]{MoE-\\BERT-Large} & SQuAD & F1 $\uparrow$ & \textbf{90.82} & 85.41 & 88.29 & 89.17\\
\hline
\multirow{2}{*}{MoE-GPT2} & \multirow{2}{*}{SAMSum} & \multirow{2}{*}{ROUGE-1 $\uparrow$} & \multirow{2}{*}{\textbf{45.56}} & \multirow{2}{*}{38.14} & \multirow{2}{*}{43.86} & \multirow{2}{*}{43.57}\\
& & & & & \\
\hline
\end{tabular}
}
\caption{The impact of token condensation on test accuracy.}
\label{tab:acc}
\end{table}
%%%
We study the impact of token condensation on model quality by evaluating three models across different datasets and metrics. Specifically, the MoE-TransformerXL model is evaluated on the WikiText-103 dataset~\cite{merity2022pointer} using the perplexity (PPL) metric, where lower PPL values indicate better performance. For MoE-BERT-Large and MoE-GPT2, we evaluate them on the SQuAD~\cite{rajpurkar2016squad} and SAMSum~\cite{gliwa2019samsum} datasets, using F1 and ROUGE-1 metrics, respectively. The larger F1 and ROUGE-1 values indicate better performance. The results are shown in \autoref{tab:acc}. When a static threshold of 0.3 is applied, the test accuracy experiences a significant drop. For example, the F1 score of the MoE-BERT-Large model decreases from 90.82 to 85.41 under a threshold of 0.3. In contrast, our proposed \textsc{Luffy} model, employing an adaptive condensation strategy, preserves the model's performance while delivering a significant training speedup.

% \subsection{\textsc{Luffy} Overhead}
% %% convergence
% \begin{figure*}[t]
% \centering
%         \subfigure[MoE-TransformerXL.]{
% 		\begin{minipage}[b]{0.31\textwidth}
% 		\includegraphics[width=1\textwidth]{Figures/Experiment/addition_overhead_xl.eps} 
%             \end{minipage}
% 		\label{fig:add_overhead_xl}
% 	}
%         \subfigure[MoE-BERT-Large.]{
%             \begin{minipage}[b]{0.31\textwidth}
%             \includegraphics[width=1\textwidth]{Figures/Experiment/addition_overhead_bert.eps}
%             \end{minipage}
%         \label{fig:add_overhead_bert}
%         }
%         \subfigure[MoE-GPT2.]{
%             \begin{minipage}[b]{0.31\textwidth}
%             \includegraphics[width=1\textwidth]{Figures/Experiment/addition_overhead_gpt.eps}
%             \end{minipage}
%         \label{fig:add_overhead_gpt}
%         }
% 	\caption{\label{fig:add_overhead}Additional overhead introduced by \textsc{Luffy}.}
% \end{figure*}
% %%%

%%%
\begin{figure*}[t]
\centering
        \subfigure[Impact of different candidate sizes.]{
		\begin{minipage}[b]{0.225\textwidth}
			\includegraphics[width=1\textwidth]{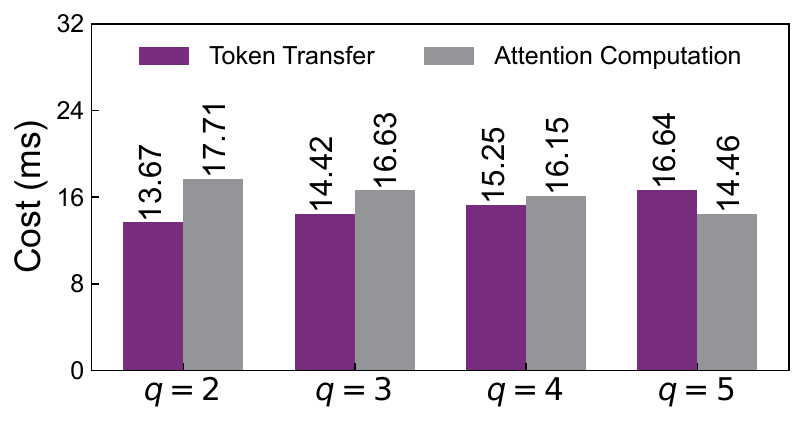} 
  \end{minipage}
		\label{fig:candidate}
	}
        \subfigure[Accuracy of the cost model.]{
            \begin{minipage}[b]{0.225\textwidth}
            \includegraphics[width=1\textwidth]{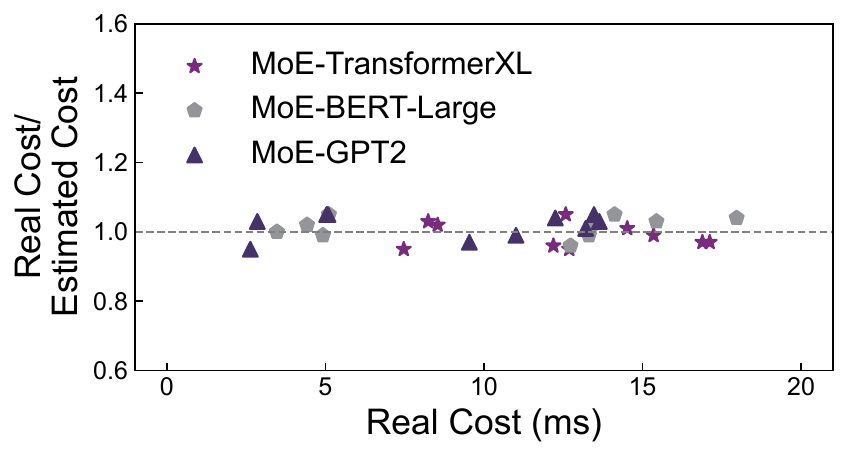}
            \end{minipage}
        \label{fig:cost_model}
        }
        \subfigure[Impact of the fast similarity measurement.]{
            \begin{minipage}[b]{0.225\textwidth}
            \includegraphics[width=1\textwidth]{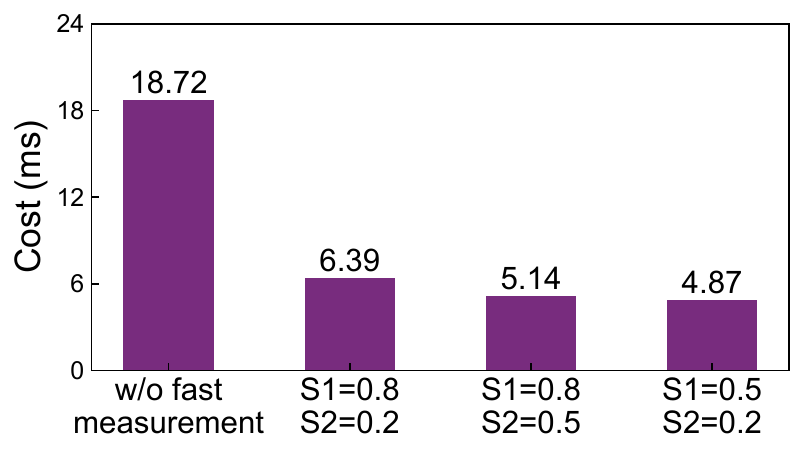}
            \end{minipage}
        \label{fig:fast_cost}
        }
        \subfigure[Training convergence with different configurations of fast similarity measurement.]{
            \begin{minipage}[b]{0.225\textwidth}
            \includegraphics[width=1\textwidth]{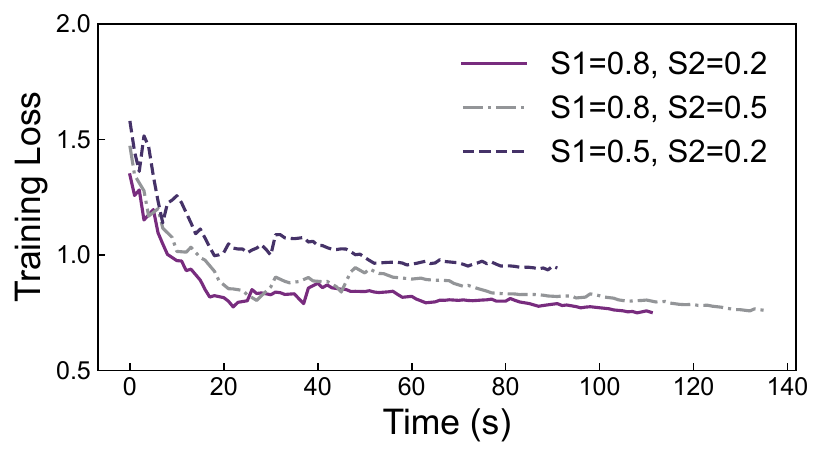}
            \end{minipage}
        \label{fig:fast_loss}
        }
	\caption{\label{fig:sen_eva}Sensitivity analysis on migration algorithm and fast similarity measurement.}
\end{figure*}
%%%

% We analyze the additional overhead introduced by \textsc{Luffy}, including sequence migration and token condensation overhead, as shown in \autoref{fig:add_overhead}. We report the overhead as a percentage of the total training time. 
% We can see that \textsc{Luffy} has trivial overhead, about only 5\% across three models. 
% In addition, we find that the sequence migration and token condensation overheads show different trends with more experts. 
% %In particular, as the number of experts increases, the sequence migration overhead increases while the token filtering overhead decreases. 
% When there are more experts, the complexity of the sequence migration algorithm increases, resulting in higher overhead. On the other hand, as more experts and GPUs are used, each GPU is assigned fewer sequences because the batch size is constant. Thus, each GPU has fewer tokens to measure similarity in the token condensation module, resulting in lower overhead. 

\subsection{Sensitivity Analysis}
% \begin{figure*}[t]
% \centering
%         \subfigure[Impact of different candidate sizes.]{
% 		\begin{minipage}[b]{0.225\textwidth}
% 			\includegraphics[width=1\textwidth]{Figures/Experiment/candidates.eps} 
%   \end{minipage}
% 		\label{fig:candidate}
% 	}
%         \subfigure[Accuracy of the cost model.]{
%             \begin{minipage}[b]{0.225\textwidth}
%             \includegraphics[width=1\textwidth]{Figures/Experiment/performance_model.eps}
%             \end{minipage}
%         \label{fig:cost_model}
%         }
%         \subfigure[Impact of the fast similarity measurement.]{
%             \begin{minipage}[b]{0.225\textwidth}
%             \includegraphics[width=1\textwidth]{Figures/Experiment/performance_fast_measurement_cost.eps}
%             \end{minipage}
%         \label{fig:fast_cost}
%         }
%         \subfigure[Training convergence with different configurations of fast similarity measurement.]{
%             \begin{minipage}[b]{0.225\textwidth}
%             \includegraphics[width=1\textwidth]{Figures/Experiment/performance_fast_measurement_loss.eps}
%             \end{minipage}
%         \label{fig:fast_loss}
%         }
% 	\caption{\label{fig:sen_eva}Sensitivity analysis on migration algorithm and fast similarity measurement.}
% \end{figure*}

We study the sensitivity of system parameters used in the sequence migration algorithm (\autoref{sec:migration_algorithm}) and the fast similarity measurement (\autoref{sec:fast_filtering}). We use the MoE-TransformerXL model to conduct the evaluation.

\noindent\textbf{Parameters of migration algorithm.} In the first step of sequence migration algorithm, top-$q$ GPUs with minimum traffic are selected. We change the value of $q$ and show the corresponding traffic and computation time in \autoref{fig:candidate}.
We can see that more candidate GPUs can reduce the attention computation cost since each sequence has more choices to stay with others of similar lengths. In contrast, a small candidate size means we mainly focus on traffic optimization, and the cost of token transfer is minimized.

We also evaluate the effectiveness of the cost model of attention computation (\autoref{sec:performance_modeling}). We collect the real costs of attention computation under different data inputs, e.g., number of sequences and sequence lengths. 
%The cost is for a single operation, not for the whole model. 
We compare the estimated cost with the real cost and report the results in \autoref{fig:cost_model}. It can be observed that our performance model introduces only a trivial error in the estimation of computation cost, with an average error of about 5\% across all models.

\noindent\textbf{Parameters of fast similarity measurement.} We study the impact of the fast similarity measurement by setting different configurations of $S_{1}$ and $S_{2}$. First, we study the impact of these parameters on the cost of similarity measurement. As shown in \autoref{fig:fast_cost}, we can see that the measurement cost can be significantly reduced when $S_{1}$ and $S_{2}$ become close, because the similarity values of less pairs need to be re-calculated.
%For example, with a setting of $S_{1}=0.8$ and $S_{2}=0.2$, the cost for similarity measurement reduced by 2.93$\times$. Furthermore, both decreasing $S_{1}$ and increasing $S_{2}$ can further reduce the cost for similarity measurement since more tokens will be directly regarded as similar or dissimilar without similarity measurement. 

We then study the impact of $S_{1}$ and $S_{2}$ on the effectiveness of the convergence. The training loss over time is shown in \autoref{fig:fast_loss}. 
When we increase $S_{2}$, more token pairs are directly regarded as dissimilar and assigned with a similarity of 0. In other words, fewer tokens are condensed in the dispatch and combine phases, and the training time is prolonged. 
In contrast, decreasing $S_{1}$ makes more token pairs be assigned with a similarity of 1, indicating that more tokens can be condensed and the total training time is reduced. However, some token pairs may be wrongly estimated as similar, which brings a negative impact to the training convergence.

%-------------------------------------------------------------------------------
\section{Related Work}\label{sec:rw}
%-------------------------------------------------------------------------------
\noindent\textbf{MoE Models.} Existing works show that model quality is strongly associated with the number of model parameters~\cite{devlin2018bert, radford2019language,wang2022language}. Recently, MoE has been widely applied as a promising solution to increase the model size and improve the model quality~\cite{qin2020multitask, zuo2021taming, dai2021generalizable, bao2022vlmo, fedus2022switch}. PaLM~\cite{chowdhery2023palm} and GLaM~\cite{du2022glam}, proposed by Google, achieve surprising results in various language tasks, such as language modeling and machine translation. Recently, Mixtral 8$\times$7B~\cite{jiang2024mixtral} has been released by Mistral AI, which achieve near state-of-the-art performance on various tasks. The success of this model inspires severl follow-up works, such as LLaMA-MoE~\cite{team2023llama}, OpenMoE~\cite{xue2024openmoe}, and DeepSeekMoE~\cite{dai2024deepseekmoe}.

\noindent\textbf{Distributed MoE Training.} The MoE models have giant model sizes and are typically trained with multiple GPUs, using an expert parallelism~\cite{xu2021gspmd, he2021fastmoe, hwang2023tutel}. Existing works introduce a series of techniques to optimize the efficiency of distributed MoE training. BASE layers~\cite{lewis2021base} implements expert parallelism based on FairSeq~\cite{ott2019fairseq}. DeepSpeed-MoE~\cite{rajbhandari2022deepspeed} introduces a hierarchical all-to-all algorithm to reduce communication costs. Tutel~\cite{hwang2023tutel} introduces the switchable parallelism and dynamic pipeline to handle unbalanced workloads of MoE. Followed by this work, PipeMoE~\cite{shi2023pipemoe} and MPipeMoE~\cite{zhang2023mpipemoe} study adaptive technologies to find optimal pipeline settings to improve the efficiency of pipeline parallelism for MoE training. SE-MoE~\cite{shen2022se} also adopts a hierarchical all-to-all algorithm to improve communication efficiency. Alpa~\cite{zheng2022alpa} develops the automated parallelism for MoE models, considering both inter-operator and intra-operator parallelism. SmartMoE~\cite{zhai2023smartmoe} studies automated parallelism and decomposes the search space into static pools for efficient hybrid parallelism searching. Lina~\cite{li2023accelerating} systematically analyzes all-to-all overhead and designs a novel communication scheduling scheme to improve all-to-all efficiency. ScheMoE~\cite{shi2024schemoe} introduces a framework to schedule communication and computation tasks in MoE training. However, these existing works cannot reduce the data transmission size for token push and pull operations, which is the main bottleneck for distributed MoE training. \textsc{Luffy} introduces two novel techniques to significantly reduce the total data transmission, improving the MoE training efficiency. 

Janus~\cite{liu2023janus} adopts a data-centric paradigm to reduce communication costs by transferring experts, which typically have smaller sizes than tokens. FasterMoE~\cite{he2022fastermoe} introduces a dynamic shadowing approach, which only transfers popular experts instead of tokens among GPUs, to reduce communication costs and achieve workload balance. Although Janus and FasterMoE can effectively reduce data transmission, they introduce intensive GPU resource competition and reduce the parallelism levels of expert running. In contrast, \textsc{Luffy} reduces the data transmission by migrating sequences, instead of transferring experts, which always parallelizes expert running at the maximum level.

\section{Conclusion}\label{sec:conc}
We present \textsc{Luffy}, an efficient distributed MoE training system. \textsc{Luffy} jointly optimizes the communication and computation efficiencies for MoE training via two novel designs. First, \textsc{Luffy} migrates sequences among GPUs to reduce the total data transmissions in the token combine phase. Second, we observe that there is a considerable number of tokens pushed to the same expert are similar, and we propose a token condensation technique to condense similar tokens in the dispatch phase, further reducing inter-GPU traffic. We implement \textsc{Luffy} and perform extensive evaluation to show that \textsc{Luffy} can significantly improve MoE training efficiency.

\bibliographystyle{IEEEtran}
\bibliography{IEEEabrv,ref}

\end{document}